\input amssym.def
\input amssym
\documentstyle{article}
\newfont{\tenbi}{cmbxti10}
\@addtoreset{figure}{section}
\def\thefigure{\thesection.\@arabic\c@figure}
\def\fps@figure{h, t}
\@addtoreset{equation}{section}

\begin{document}
\newtheorem{thm}{Theorem}[section]
\newtheorem{prop}[thm]{Proposition}
\newtheorem{lem}[thm]{Lemma}
\newtheorem{cor}[thm]{Corollary}

\title{Dissipation Induced Instabilities}

\author{
Anthony Bloch
   \thanks{Research partially supported by the National Science
        Foundation grant DMS--90--02136, PYI grant DMS--91--57556, and AFOSR
        grant F49620--93--1--0037}
   \\Department of Mathematics, Ohio State University
   \\Columbus, OH 43210  \and
P.S.Krishnaprasad
   \thanks{Research partially supported by the AFOSR University Research
        Initiative Program under grants AFOSR-87-0073 and AFOSR-90-0105 and
by the National
        Science Foundation's Engineering Research Centers Program NSFD CDR
8803012}
   \\Department of Electrical  Engineering and
   \\Institute for Systems Research
   \\ University of Maryland, College Park, MD 20742
\and
Jerrold E. Marsden
   \thanks{Research partially supported by, DOE contract
DE--FG03--92ER--25129, a
        Fairchild Fellowship at Caltech, and the Fields Institute for
Research in the
        Mathematical Sciences}
   \\Department of Mathematics
   \\University of California,  Berkeley, CA 94720
\and
Tudor S. Ratiu
   \thanks{Research partially supported by NSF Grant DMS 91--42613 and DOE
contract
        DE--FG03--92ER--25129}
   \\Department of Mathematics
   \\University of California,  Santa Cruz, CA 95064}
\date{August, 1990; this version, March 9, 1993}
\maketitle

\begin{abstract}
The main goal of this paper is to prove that if the energy-momentum
(or energy-Casimir) method predicts formal instability of a relative
equilibrium in
a Hamiltonian system with symmetry, then with the addition of dissipation, the
relative equilibrium becomes spectrally and hence linearly and nonlinearly
unstable.
The energy-momentum method assumes that one is in the context of a
mechanical system
with a given symmetry group. Our result assumes that the dissipation chosen
does not
destroy the conservation law associated with the given symmetry group---thus,
we
consider internal dissipation. This also includes the special case of systems
with no symmetry and ordinary equilibria. The theorem is proved by combining
the
techniques of Chetaev, who proved instability theorems using a special
Chetaev-Lyapunov function, with those of Hahn, which enable one to
strengthen the
Chetaev results from Lyapunov instability to spectral instability. The main
achievement
is to strengthen Chetaev's methods to the context of the block
diagonalization version of
the energy momentum method given by Lewis, Marsden, Posbergh, and Simo.
However, we also
give the eigenvalue movement formulae of Krein, MacKay and others both in
general and
adapted to the context of the normal form of the linearized equations given
by the
block diagoanl form, as provided by the energy-momentum method.  A number
of specific
examples, such as the rigid body with internal rotors, are provided to
illustrate the
results.   \end{abstract}

\section{Introduction}

A central and time honored problem in mechanics is the determination of the
stability
of equilibria and relative equilibria of Hamiltonian systems. Of particular
interest
are the relative equilibria of simple mechanical systems with symmetry, that
is,
Lagrangian or Hamiltonian systems with energy of the form kinetic plus
potential
energy, and that are invariant under the canonical action of a group. {\it
Relative
equilibria\/} of such systems are solutions whose dynamic orbit coincides
with a one
parameter group orbit. When there is no group present, we have an
equilibrium in the
usual sense with zero velocity; a relative equilibrium, however can have
nonzero
velocity.  When the group is the rotation group, a relative equilibrium is
a {\it
uniformly rotating state}.

The analysis of the stability of relative equilibria has a distinguished
history and
includes the stability of a rigid body rotating about one of its principal
axes, the
stability of rotating gravitating fluid masses and other rotating systems.
(See for
example, Riemann [1860], Routh [1877], Poincar\'{e} [1885, 1892, 1901], and
Chandrasekhar [1977]).

Recently, two distinct but related systematic methods have been developed
to analyze
the stability of the relative equilibria of Hamiltonian systems. The first,
the {\it
energy-Casimir\/} method, goes back to Arnold [1966] and was developed and
formalized
in Holm, Marsden, Ratiu, and Weinstein [1985], Krishnaprasad and Marsden
[1987] and
related papers. While the analysis in this method often takes place in a linear
Poisson reduced space, often the ``body frame'', and this is sometimes
convenient, the
method has a serious defect in that a lack of sufficient Casimir functions
makes it
inapplicable to examples such as geometrically exact rods, three
dimensional ideal
fluid mechanics, and some plasma systems.

This deficiency was overcome in a series of papers developing and applying the
{\it energy-momentum\/} method; see Marsden, Simo, Lewis and Posbergh
[1989], Simo,
Posbergh and Marsden [1990, 1991], Lewis and Simo [1990], Simo, Lewis, and
Marsden
[1991], Lewis [1992], and Wang and Krishnaprasad [1992]. These techniques
are based on
the use of the Hamiltonian plus a conserved quantity. In the
energy-momentum method,
the relevant combination is the {\it augmented Hamiltonian\/}. One can
think of the
energy momentum method as a synthesis of the ideas of Arnold for the group
variables,
and those of Routh and Smale for the internal variables. In fact, one of
the bonuses
of the method is the appearance of normal forms for the energy and the
symplectic
structure, which makes the method particularly powerful in applications.

The above techniques are designed for conservative systems. For these systems,
but especially for dissipative systems, {\it spectral methods\/} pioneered
by Lyapunov
have also been powerful. In what follows, we shall elaborate on the
relation with the
above energy methods.

The key question that we address in this paper is: if the energy momentum
method
predicts formal instability, {\it i.e.,\/} if the augmented energy has a
critical
point at which the second variation is not positive definite, is the system
in some
sense {\it unstable\/}? Such a result would demonstrate that the
energy-momentum
method gives sharp results. The main result of this paper is that this is
indeed true
if small dissipation, arising from a Rayleigh dissipation
function, is added to the internal variables of a system. (Dissipation in the
rotational variables will be considered in another publication.) In other
words, we
prove that   \begin{quote}  {\it If a relative equilibrium of a Hamiltonian
system with
symmetry is formally unstable by the energy-momentum method, then it is
linearly and
nonlinearly unstable when a small amount of damping (dissipation) is added
to the
system.}  \end{quote}

Some special cases of, and commentaries about, the topics of the present
paper were
previously known. As we shall discuss below, one of the main early
references for
this topic is Chetaev [1961] and some results were already known to Thomson
and Tait
[1912]. See also Ziegler [1956], Haller [1992], and Sri Namachchivaya and
Ariaratnam
[1985]. The latter paper shows the effect of dissipation induced
instabilities for
rotating shafts, and contains a number of other interesting references.
\vspace{0.2in}

Next, we outline how the program of the present paper is carried out. To do
so, we
first look at the case of ordinary equilibria. Specifically, consider an
equilibrium
point $z _e$ of a Hamiltonian vector field $X _H$ on a symplectic manifold
$P$, so
that $ X _H (z _e) = 0$ and $H$ has a critical point at $ z _e $. Then the two
standard methodologies for studying stability mentioned above are as follows:
\begin{enumerate}
\begin{enumerate}  \item {\it Energetics\/} --- determine if  \[
\delta ^2 H (z _e) =: {\cal Q}  \]  is a definite quadratic form (the
Lagrange-Dirichlet criterion). \item {\it Spectral methods\/} --- determine
if the
spectrum of the linearized operator \[{\bf D} X _H (z _e) =: L \]
is on the imaginary axis.
\end{enumerate}
\end{enumerate}

The energetics method can, via ideas from reduction, be applied to relative
equilibria too and this is the basis of the energy-momentum method alluded
to above and
which we shall detail in \S {\bf 3}. The spectral method can also be applied to
relative equilibria since under reduction, a relative equilibrium becomes an
equilibrium.

For general (not necessarily Hamiltonian) vector fields, the classical
Lyapunov theorem states that if the spectrum of the linearized equations
lies strictly
in the left half plane, then the equilibrium is stable and even
asymptotically stable
(trajectories starting close to the equilibrium converge to it
exponentially as $ t
\rightarrow \infty $). Also, if any eigenvalue is in the strict right half
plane, the
equilibrium is unstable. This result, however, cannot apply to the purely
Hamiltonian
case since the spectrum of $L$ is invariant under reflection in the real
and imaginary
coordinate axes. Thus, the only possible spectral configuration for a
stable point of
a Hamiltonian system is if the spectrum is {\it on\/} the imaginary axis.

The relation between (a) and (b) is, in general, complicated, but one can
make some
useful elementary observations.
\vspace{0.2in}

\noindent {\large \bf Remarks } \quad  \\
\noindent {\large \bf 1} \quad Definiteness of ${\cal Q}$ implies spectral
stability
({\it i.e.,\/} the spectrum of $L$ is on the imaginary axis). This is
because spectral
instability implies (linear and nonlinear) instability (Lyapunov's
Theorem), while
definiteness of ${\cal Q}$ implies stability (the Lagrange Dirichlet
criterion).

\noindent {\large \bf 2} \quad Spectral stability need not imply stability,
even
linear stability. This is shown by the unstable linear system $ \dot{q} =
p, \dot{ p} =
0 $ with a pair of eigenvalues at zero. Other resonant examples exhibit similar
phenomena with nonzero eigenvalues.

\noindent {\large \bf 3} \quad If ${\cal Q}$ has odd index (an odd number
of negative
eigenvalues), then $L$ has a real positive eigenvalue. This is a special case
of
theorems of Chetaev [1961] and Oh [1987]. Indeed, in canonical coordinates, and
identifying ${\cal Q}$ with its corresponding matrix, we have  \[ L = {\Bbb
J} {\cal
Q}. \] Thus, $ \det L = \det {\cal Q} $ is negative. Since $ \det L $ is
the product of
the eigenvalues of $ L $ and they come in conjugate pairs, there must be at
least one
pair of real eigenvalues, and since the set of eigenvalues is invariant under
reflection in the imaginary axis, there must be an odd number of positive real
eigenvalues.

\noindent {\large \bf 4} \quad If $ P = T ^\ast Q $ with the standard cotangent
symplectic structure and if $H$ is of the form kinetic plus potential so that
an
equilibrium has the form $ (q _e, 0) $, and if $ \delta ^2 V (q _e) $ has
nonzero
(even or odd) index, then again $L$ must have real eigenvalues. This is
because one
can diagonalize $ \delta ^2 V (q _e) $ with respect to the kinetic energy inner
product, in which case the eigenvalues are evident. In this context, note
that there
are no gyroscopic forces. \quad $\blacklozenge$  \vspace{0.2in}

To get more interesting effects than covered by the above examples, we consider
{\tenbi gyroscopic systems\/}; {\it i.e.,\/} linear systems of the form
\begin{equation}\label{1071}  M \ddot{q} + S \dot{ q} + \Lambda  q = 0
\end{equation}
where $ q \in {\Bbb R} ^n $,  $M$ is a positive definite symmetric $ n
\times n $
matrix, $S$ is skew, and $ \Lambda $ is symmetric. This system is verified to
be
Hamiltonian with $ p = M \dot{q} $, energy function
\begin{equation}\label{1072}  H
(q,p) = \frac{1}{2} p M ^{-1} p + \frac{1}{2} q \Lambda q  \end{equation}
and the bracket
\begin{equation}\label{1073}
\{ F, K \} = \frac{\partial F}{\partial q ^i} \frac{\partial K}{\partial p_i} -
\frac{\partial K}{\partial q^i} \frac{\partial F}{\partial p_i} - S_{ij}
\frac{\partial F}{\partial p_i} \frac{\partial K}{\partial p_j}.
\end{equation}
Systems of this form arise from simple mechanical systems via reduction;
this form is in
fact the normal form of the linearized equations when one has an {\it
abelian\/} group. Of
course, one can also consider linear systems of this type when gyroscopic
forces are added
ab initio, rather than being derived by reduction. Such systems arise in
control theory,
for example; see Bloch, Krishnaprasad, Marsden, and Sanchez [1991] and Wang and
Krishnaprasad [1992].

If the index of $V$ is even (see Remark {\bf 3}) one can get situations
where $ \delta ^2
H $ is indefinite and yet spectrally stable. Roughly, this is a situation
that is
capable of undergoing a Hamiltonian Hopf bifurcation. One of the simplest
systems in which
this occurs is in the linearized equations about a special relative
equilibrium, called
the ``cowboy'' solution, of the double spherical pendulum; see Marsden and
Scheurle [1992]
and \S {\bf 6} below. Another example arises from certain solutions of the
heavy top
equations as studied in Lewis, Ratiu, Simo and Marsden [1992]. Other
examples are
given in \S {\bf 6}. One of our first main results is the following:

\begin{thm} {\bf Dissipation induced instabilities---abelian case\/} Under
the above
conditions, if we modify~(\ref{1071}) to
\begin{equation}\label{1074}
M \ddot{ q} + (S + \epsilon R) \dot{q} + \Lambda q = 0
\end{equation}
for small $ \epsilon > 0 $, where $R$ is symmetric and positive definite,
then the
perturbed linearized equations  \[ \dot{z} = L _\epsilon z, \]
where $ z = (q, p)$ are spectrally unstable, {\it i.e.,\/} at
least one pair of eigenvalues of $ L _\epsilon $ is in the right half plane.
\end{thm}

This result builds on basic work of Thomson and Tait [1912],
Chetaev [1961], and Hahn [1967]. The argument proceeds in two steps.
\vspace{0.2in}

\noindent {\bf Step 1\/} {\it Construct the\/} {\tenbi Chetaev function\/}
\begin{equation}\label{1075}
W (q,p) = H (q,p) + \beta M ^{-1} p \cdot (\Lambda q)
\end{equation}
{\it for small\/} $\beta$ and use this to prove Lyapunov instability.
\vspace{0.2in}

This function has the key property that for $\beta$ small enough, $W$ has the
same index as $H$, yet $ \dot{ W} $ is {\it negative definite\/}, where the
overdot is
taken in the dynamics of~(\ref{1074}). This is enough to prove Lyapunov
instability, as is seen by studying the equation
\begin{equation}\label{1076}
W (q (T), p (T)) = W (q_0 ,p_0) + \int^T_0 \dot{ W} (q (t), p (t)) d t
\end{equation}
and choosing $ (q _0, p _0) $ in the sector where $W$ is negative, but
arbitrarily close to the origin.
\vspace{0.2in}

\noindent {\bf Step 2\/} {\it Employ an argument of Hahn\/} [1967] {\it to show
spectral instability\/}. \vspace{0.2in}

The sketch of the proof of step 2 is as follows. Since $\epsilon$ is small
and the original
system is Hamiltonian, the only nontrivial possibility to exclude is the case
in which the unperturbed eigenvalues lie on the imaginary axis at nonzero
values and that,
after perturbation, they remain on the imaginary axis. Indeed, they
cannot all move left by Step 1 and $ L _\epsilon $ cannot have zero
eigenvalues since $ L
_\epsilon z = 0 $ implies $ \dot{ W} (z,z) = 0 $. However, in this case,
Hahn [1967] shows
the existence of at least one periodic orbit, which cannot exist in view
of~(\ref{1076})
and the fact that $ \dot{ W} $ is negative definite. The details of these
two steps are carried out in \S{\bf 3} and \S{\bf 4}.
\vspace{0.2in}

This theorem generalizes in two significant ways. First, it is valid for
infinite
dimensional systems, where $ M, S, R $ and $ \Lambda  $ are replaced by linear
operators. One of course needs some technical conditions to ensure that $W$
has the
requisite properties and that the evolution equations generate a semi-group
on an
appropriate Banach space. For Step 2 one requires, for example, that the
spectrum at
$\epsilon = 0$ be discrete with all eigenvalues having finite multiplicity.
To apply
this to nonlinear systems under linearization, one also needs to know that the
nonlinear system satisfies some ``principle of linearized stability''; for
example, it
has a good invariant manifold theory associated with it.

The second generalization is to systems in block diagonal form but with a
non-abelian group. The system~(\ref{1074}) is the form that block
diagonalization
gives with an abelian symmetry group. For a non-abelian group, one gets,
roughly
speaking, a system consisting of~(\ref{1074}) coupled with a Lie-Poisson
(generalized
rigid body) system. The main step needed in this case is a significant
generalization
of the Chetaev function. This is carried out in \S {\bf 3}.

A nonabelian example (with the group $SO(3)$) that we
consider in \S {\bf 6} is the rigid body with
internal momentum wheels.

 The formulation of Theorem {\bf 1.1} and its generalizations is attractive
because of
the interesting conclusions that can be obtained essentially from
energetics alone. If
one is willing to make {\it additional\/} assumptions, then there is a
formula giving
the amount by which {\it simple\/} eigenvalues move {\it off\/} the
imaginary axis.
One version of this formula, due to MacKay [1991], states that\footnote{As
Mark Levi
has pointed out to us, formulae like (1.7) go back to Krein [1950] and Krein
also obtained such formulae for periodic orbits (see Levi [1977], formula
(18), p.
33).}  \begin{equation}\label{1077}  {\rm Re}\, \lambda _\epsilon = \frac{
\bar{ \xi}
({\Bbb J} B)_{\rm anti} \xi }{\bar{ \xi} ^T {\Bbb J} \xi} \epsilon + O
(\epsilon ^2)
\end{equation}  where we write the linearized equations in the form
\begin{equation}\label{1078}  \dot{z} = L _\epsilon z = ({\Bbb J} {\cal Q}
+ \epsilon
B) z . \end{equation}  Here, $ \lambda _\epsilon $ is the perturbed eigenvalue
associated with a simple eigenvalue $ \lambda _0 = i \omega _0 $ on the
imaginary axis
at $ \epsilon = 0 $, $\xi$ is a (complex) eigenvector for $ L _0 $ with
eigenvalue $
\lambda _0 $, and $ ({\Bbb J} B)_{\rm anti} $ is the antisymmetric part of
$ {\Bbb J}
B $. \vspace{0.2in}

In fact, the ratio of quadratic functions in~(\ref{1077}) can be replaced by a
ratio involving energy-like functions and their time derivatives including
the energy
itself or the Chetaev function. To actually work out~(\ref{1077}) for examples
like~(\ref{1071}) can involve considerable calculation. See \S {\bf 5} for
details.
\vspace{0.2in}

What follows is a simple example in which one can carry out the analysis to a
large extent directly. We hasten to add that problems like the double spherical
pendulum are considerably more complex algebraically and a direct analysis
of the
eigenvalue movement would not be so simple.

Consider the following gyroscopic system
(cf. Chetaev [1961] and Baillieul and Levi [1991])
\begin{equation}\label{1079}
\left. \begin{array}{cc}
\ddot{ x} - g \dot{y} + \gamma \dot{x}
            + \alpha x  =   0 \\ \mbox{  } \\
\ddot{ y} + g \dot{x} + \delta \dot{y} + \beta
           y  =  0 \end{array} \right\}
\end{equation}
which is a special case of~(\ref{1074}). Assume $ \gamma \geq 0 $ and $ \delta
\geq 0 $. For $ \gamma = \delta = 0 $ this system is Hamiltonian with
symplectic
form \begin{equation}\label{10710}
\Omega = d x \wedge d \dot{x} + d y \wedge d \dot{y} - g d x \wedge d y
\end{equation}
and the bracket (1.3) where $ S = \left( \begin{array}{cc} 0 & - g \\ g & 0
\end{array} \right)
$ and Hamiltonian
\begin{equation}\label{10711}
H = \frac{1}{2} (\dot{x} ^2 + \dot{y} ^2) + \frac{1}{2} (\alpha x^2 + \beta
y^2).
\end{equation}
Note that for $ \alpha = \beta $, angular momentum is conserved
corresponding to the
$S^1$ symmetry of $H$. The characteristic polynomial is computed to be
\begin{equation}\label{10712}  p (\lambda) = \lambda ^4 + (\gamma + \delta)
\lambda ^3
+ ( g ^2 + \alpha + \beta + \gamma \delta) \lambda ^2 + (\gamma \beta +
\delta \alpha)
\lambda + \alpha \beta . \end{equation} Let the characteristic polynomial
for the
undamped system be denoted $ p _0 $: \begin{equation}\label{10713}  p _0
(\lambda) =
\lambda ^4 + (g ^2 + \alpha + \beta) \lambda ^2 + \alpha \beta. \end{equation}
Since
$ p _0 $ is quadratic in $ \lambda ^2 $, its roots are easily found. One gets:

\begin{description}
\item[\ \ {\bf i}] \mbox{ } If $ \alpha > 0, \beta > 0$, then $ H $ is positive
definite and the eigenvalues are on the imaginary axis; they are coincident
in a 1 : 1
resonance for $ \alpha = \beta $.
\item[\ {\bf ii}] \mbox{ \ } If $\alpha$ and
$\beta$ have opposite signs, then $H$ has index 1 and there is one
eigenvalue pair on
the real axis and one pair on the imaginary axis.
\item[{\bf iii}]	\mbox{\ \ } If $
\alpha < 0 $ and $ \beta < 0 $ then $H$ has index 2. Here the eigenvalues
may or may
not be on the imaginary axis.  \end{description}

To determine what happens in the last case, let
\[ D = (g ^2 + \alpha + \beta) ^2 - 4 \alpha \beta = g ^4 + 2 g ^2 (\alpha
+ \beta) +
(\alpha - \beta)^2 \]
be the discriminant, so that the roots of~(\ref{10713}) are given by
\[ \lambda ^2 = \frac{1}{2} [ - (g ^2 + \alpha + \beta) \pm \sqrt{D} ]. \]
Thus we arrive at the following conclusions:
\begin{description}
\item[\ \ {\bf a}] \mbox{ } If $ D < 0$, then there are two roots in the
right half
plane and two in the left.  \item[\ \ {\bf b}] \mbox{ } If $D = 0 $ and $ g
^2 + \alpha
+ \beta > 0 $, there are coincident roots on the imaginary axis, and if $ g ^2
+
\alpha + \beta < 0 $, there are coincident roots on the real axis. \item[\
\ {\bf c}]
\mbox{ } If $ D > 0 $ and $ g ^2 + \alpha + \beta > 0 $, the roots are on the
imaginary axis and if $ g ^2 + \alpha + \beta < 0 $, they are on the real axis.
\end{description} Thus the case in which $ D \geq 0 $ and $ g ^2 + \alpha +
\beta > 0
$ ({\it i.e.,\/} if $ g ^2 + \alpha + \beta \geq 2 \sqrt{ \alpha \beta} $),
is one to
which the dissipation induced instabilities theorem (Theorem {\bf 1.1})
applies.

Note that for $ g ^2 + \alpha + \beta > 0 $, if $D$ decreases through zero, a
Hamiltonian Hopf bifurcation occurs. For example, as $g$ increases and the
eigenvalues
move onto the imaginary axis, one speaks of the process as {\tenbi gyroscopic
stabilization\/}.

Now we add damping and get

\begin{prop} If $ \alpha < 0, \beta < 0, D > 0, g ^2 + \alpha + \beta > 0 $
and least
one of $ \gamma, \delta $ is strictly positive, then for~(\ref{1079}), there is
exactly one pair of eigenvalues in the strict right half plane. \end{prop}

\noindent {\bf Proof\,} We use the Routh-Hurwitz criterion (see Gantmacher
[1959, vol.
2]), which states that the number of strict right half plane roots of the
polynomial
\[ \lambda ^4 + \rho_1  \lambda ^3 + \rho _2 \lambda ^2 + \rho _3 \lambda +
\rho _4 \]
equals the number of sign changes in the sequence
\begin{equation}\label{10714}
 \left\{ 1, \rho _1 , \frac{ \rho _1 \rho _2 - \rho _3 }{ \rho _1}, \frac{
\rho _3
(\rho _1 \rho _2 - \rho _3) - \rho _1 \rho _4 }{ \rho _1 \rho _2 - \rho
_3}, \rho _4
\right\} .  \end{equation}   For our case, $ \rho _1 = \gamma + \delta > 0,
\rho _2 =
g ^2 + \alpha + \beta + \gamma \delta > 0, \rho _3 = \gamma \beta + \alpha
\delta < 0
$ and $ \rho _4 = \alpha \beta > 0 $, so the sign sequence~(\ref{10714}) is
\[ \{ +, +, +, -, + \}. \]
Thus, there are two roots in the right half plane. \quad $\blacksquare$
\vspace{0.2in}

This proof confirms the result of Theorem {\bf 1.1}. It gives more
information, but
for complex systems, this method, while instructive, may be difficult or
impossible to
implement, while the method of Theorem {\bf 1.1} is easy to implement. One
can also
use methods of Krein and MacKay to get the result of the above proposition
and get, in
fact, additional information about how far the eigenvalues move to the
right as a
function of the size of the dissipation. We shall present this technique in
\S {\bf
5}. Again, this technique gives more specific information, but is harder to
implement,
as it requires more hypotheses (simplicity of eigenvalues) and requires one
to compute
the corresponding eigenvector of the unperturbed system, which may not be a
simple
task. \vspace{0.2in}

\noindent{\large \bf Example} An instructive special case of the system (1.9)
is
the system of equations describing a bead in equilibrium at the center of a
rotating
circular plate driven with angular velocity $\omega$ and subject to a central
restoring force. (These equations may also be regarded as the linearized
equations of
motion for a rotating spherical pendulum in a gravitational field---see
Baillieul and
Levi [1991].) Let $x$ and $y$ denote the position of the bead in a rotating
coordinate
system fixed in the plate. The Lagrangian is then \begin{equation}
\frac{1}{2} (x _t
- \omega y) ^2 + (y _t + \omega x) ^2 - \frac{1}{2} k (x^2 + y^2)
\end{equation}
and the equations of motion without damping are
\begin{equation}
\begin{array}{c}
x _{ tt} - 2 \omega y _t + (k - \omega ^2) x = 0 \\
y _{ tt} + 2 \omega x _t + (k - \omega ^2) x = 0.
\end{array}
\end{equation}
Thus for $ \omega ^2 > k $ the system is gyroscopically stable and the
addition of
Rayleigh damping induces spectral instability. \quad $\blacklozenge$
\vspace{0.2in}

It is interesting to speculate on the effect of damping on the Hamiltonian Hopf
bifurcation in view of these general results and in particular, this example.

For instance, suppose $ g ^2 + \alpha + \beta > 0 $ and we allow $D$ to
increase so
a Hamiltonian Hopf bifurcation occurs in the undamped system. Then the
above sign
sequence does not change, so no bifurcation occurs in the damped system;
the system is
unstable and the Hamiltonian Hopf bifurcation just enhances this
instability. However,
if we simulate forcing or control by allowing one of $\gamma$ or $\delta$ to be
negative, but still small, then the sign sequence is more complex and one
can get, for
example, the Hamiltonian Hopf bifurcation breaking up into two nearly
coincident Hopf
bifurcations. These remarks are consistent with van Gils, Krupa, and
Langford [1990].

The preceding discussion assumes that the equilibrium of the original
nonlinear equation being linearized is independent of $ \epsilon. $  In
general of course this is not true, but it can be dealt with as follows.
Consider the nonlinear equation
\begin{equation}
\dot{x} = f(x, \epsilon)
\end{equation}
on a Banach space, say.  Assume $ f(0, 0) = 0 $ and $ x(\epsilon) $ is a curve
of equilibria with $ x(0) = 0. $ By implicitly differentiating $
f(x(\epsilon), \epsilon) = 0 $ we find that the linearized equations at
$x(\epsilon)$ are given by
\begin{eqnarray}
\dot{ (\delta  x)} & = & {\bf D}_xf(x(\epsilon), \epsilon) \delta  x  \nonumber
\\ & = & {\bf D}_xf (0, 0) \delta  x + \epsilon \left[ {\bf D}^2_{x \epsilon}
f(0, 0) \delta x + {\bf D}^2_x f(0, 0) (x^\prime (0), \delta  x) \right]
\nonumber \\ & & + O(\epsilon^2)
\end{eqnarray}
where $ x^{\prime} (0) = - {\bf D}_x f(0, 0) ^{-1} f _\epsilon (0, 0), $
assuming
that $ {\bf D}_x f(0, 0) $ is invertible; {\it i.e., we are not at a
bifurcation
point}. In principle then, (1.17) is computable in terms of data at $ (0, 0)$
and our general theory applies.

A situation of interest for KAM theory is the study of the dynamics
near an elliptic fixed point of a Hamiltonian system with several degrees
of freedom.
The usual hypothesis is that the equations linearized about this fixed
point have a
spectrum that lies on the imaginary axis and that the second variation of the
Hamiltonian at this fixed point is indefinite. Our result says that these
elliptic fixed
points become spectrally unstable with the addition of (small) damping. It
would be of
interest to investigate the role of our result, and associated system
symmetry breaking
results (see, for example, Guckenheimer and Mahalov [1992]), for these
systems and in the
context of Hamiltonian normal forms, more thoroughly (see, for example,
Haller [1992]). In
particular, the relation between the results here and the phenomenon of
capture into
resonance would be of considerable interest.

There are a number of other topics that should be investigated in the
future. For
example, the present results would be interesting to apply to some fluid
systems.
The cases of interest here, in which eigenvalues lie on the imaginary axis, but
the second variation of the relevant energy quantity is indefinite, occur
for circular
rotating liquid drops (Lewis, Marsden, and Ratiu [1987] and Lewis [1989]), for
shear flow in a stratified fluid with Richardson number between $1/4$ and 1
(Abarbanel et
al. [1986]), in plasma dynamics (Morrison and Kotschenreuther [1989],
Kandrup [1991],
and Kandrup and Morrison [1992]), and for rotating strings. In each of
these examples,
there are essential pde difficulties that need to be overcome, and we have
written the
present paper to adapt to that situation as far as possible. One infinite
dimensional
example that we consider is the case of a rotating rod in \S{\bf 6}, but it
can be treated
by essentially finite dimensional methods, and the pde difficulties we were
alluding to do
not occur. We also point out that some of the same effects as seen here are
also found in
reversible (but non-Hamiltonian) systems; see O'Reilly [1993].
\vspace{0.2in}

\noindent{\bf Acknowledgments} We thank Stuart Antman, John Baillieul,
Gy\"{o}rgy
Haller, Mark Levi, Debbie Lewis, John Maddocks, Lisheng Wang, and Steve
Wiggins for
helpful comments. In particular, we thank Debbie Lewis for helpful comments
on the
linearized equations in the block diagonalization theory. We also thank the
Fields
Institute for providing the opportunity to meet in pleasant surroundings.

\section{The Energy-Momentum Method}

Our framework for the energy-momentum method will be that of simple mechanical
systems with symmetry. We choose as the phase space $P= TQ$ or $P = T^\ast Q$,
a
tangent or cotangent bundle of a configuration space $Q$. Assume there is a
Riemannian
metric $\langle \! \langle \, , \rangle \! \rangle $ on $Q$, that a Lie
group $G$ acts
on $Q$ by isometries (and so $G$ acts symplectically on $ TQ $ by tangent
lifts and on
$T^\ast Q $ by cotangent lifts). The Lagrangian is taken to be of the form

\begin{equation}\label{111}
L(q, v) = \frac{1}{2} \| v \|_q  ^2 - V(q),
\end{equation}
or equivalently, the Hamiltonian is
\begin{equation}\label{311}
H(q, p)  =  \frac{1}{2} \| p \|^2_q +  V(q),
\end{equation}
where $ \| \cdot \|_q $ is the norm on $ T _q Q $ or the one induced on $
T^\ast_q Q$, and where
$V$ is a $G$-invariant potential.

With a slight abuse of notation, we write either $ (q, v) $ or $ v _q $ for
a vector
based at $q \in Q $ and $ z = (q, p) $ or $ z = p_q $ for a covector based
at   $q \in
Q$. The pairing between  $T^\ast_q Q$ and $T_q Q$ is written
\begin{equation}\label{312} \langle  p_q, v_q \rangle, \quad   \langle  p,
v \rangle
\quad \mbox{or} \quad \langle  (q,p), (q,v) \rangle . \end{equation}  Other
natural
pairings between spaces and their duals are also denoted $ \langle \, ,
\rangle $.

The standard momentum map for simple mechanical $G$-systems is
\begin{eqnarray} \label{313}
{\bf J}:TQ \rightarrow \frak g^{\ast}, \quad \mbox{where} \quad \langle
{\bf J} (q,v),\xi \rangle =
\langle \!  \langle  v, \xi_Q (q) \rangle \! \rangle \nonumber \\
{\rm or \quad }{\bf J} : T^\ast Q \rightarrow \frak g ^{\ast}, \quad
\mbox{where} \quad  \langle
{\bf J} (q,p), \xi \rangle = \langle
 p, \xi_Q (q) \rangle
\end{eqnarray}
where $ \xi_Q $ denotes the infinitesimal generator of $ \xi \in \frak g $
on $ Q $. We use the same
notation for {\bf J} regarded as a map on either the cotangent or the
tangent space; which is meant
will be clear from the context. For future use, we set $ \frak g \cdot q =
\{\xi _Q (q)
\mid \xi \in \frak g \} \subset T _q Q $.

Assume that $G$ acts freely on $Q$ so we can regard $Q \rightarrow Q/G$ as a
principal $G$-bundle. A refinement shows that one really only needs the
action of $ G _\mu $ on
$Q$ to be free and all the constructions can be done in terms of the bundle
$ Q \rightarrow Q/G
_\mu $; here, $G _\mu$ is the isotropy subgroup for $\mu \in \frak g
^{\ast}$ for the coadjoint
action of $G$ on $\frak g ^{\ast}$. Recall that for abelian groups, $G = G
_\mu$. However, we do
the constructions for the action of the full group $G$ for simplicity of
exposition.

For each $q \in Q$, let the {\tenbi locked inertia
tensor\/} be the map ${\Bbb I} (q): \frak g \rightarrow \frak g ^{\ast}$
defined by
\begin{equation}\label{331}
\langle  {\Bbb I} (q) \eta , \zeta \rangle = \langle \! \langle \eta_Q (q),
\zeta_Q
(q) \rangle \! \rangle .
\end{equation}
Since the action is free, ${\Bbb I} (q)$ is indeed an inner product. The
terminology
comes from the fact that for coupled rigid or elastic systems, ${\Bbb I}
(q)$ is the
classical moment of inertia tensor of the corresponding rigid system. Most
of the results of this
paper hold in the infinite as well as the finite dimensional case. To
expedite the exposition, we
give many of the formulae in coordinates for the finite dimensional case.
For instance,
\begin{equation}\label{332} {\Bbb I} _{ab}  =  g_{ij}A^i_{\;a} A^j_{\;b},
\end{equation}
where we write
\begin{equation}\label{2.7}
[\xi _Q(q)] ^i = A^i_{\;a}(q)\xi ^a
\end{equation}
relative to coordinates $ q ^i , i = 1, 2, \ldots , n $ on $Q$ and a basis
$ e _a , a = 1,
2, \ldots , m $ of $\frak g$.

Define the map $\alpha : TQ \rightarrow \frak g$ which assigns to each
$(q, v)$ the corresponding {\tenbi angular velocity of the locked system\/}:
\begin{equation}\label{333}
\alpha (q, v)  =  {\Bbb I}(q)^{-1} ({\bf J} (q, v)).
\end{equation}
In coordinates,
\begin{equation}\label{334}
\alpha^a  =  {\Bbb I}^{ab} g_{ij}A^i_{\;b}  v^j.
\end{equation}

The map~(\ref{333}) is a connection on the principal $G$-bundle $Q
\rightarrow Q/G$. In
other words, $\alpha$ is $G$-equivariant and satisfies $\alpha (\xi_Q(q)) =
\xi$, both of
which are readily verified. In checking equivariance one uses invariance of
the metric,
equivariance of $ {\bf J} : TQ \rightarrow \frak g ^{\ast}$, and
equivariance of  ${\Bbb
I}$ in the sense of a map ${\Bbb I} : Q \rightarrow {\cal L} (\frak g,
\frak g ^{\ast})$
({\it i.e.,\/} the space of linear maps of $\frak g$ to $ \frak g ^{\ast}
$), namely ${\Bbb
I} (g \cdot q) \cdot A d_g \xi = A d ^\ast _{g ^{-1}} {\Bbb I} (q) \cdot \xi$.

We call $\alpha$ the {\tenbi mechanical connection\/}, as in Simo, Lewis
and Marsden [1991].
The horizontal space of the connection $\alpha$ is given by
\begin{equation}\label{335}
{\rm hor}_q = \{(q, v) \mid {\bf J}(q, v) = 0\} ;
\end{equation}
{\it i.e.,} the space orthogonal to the $G$-orbits. The vertical space
consists of
vectors that are mapped to zero under the projection $Q \rightarrow S =
Q/G$; {\it i.e.,}
\begin{equation}\label{336}
{\rm ver}_q  =  \{\xi_Q(q) \mid \xi \in \frak g \} = \frak g \cdot q.
\end{equation}

For each $ \mu \in \frak g ^{\ast} $, define the 1-form $ \alpha_\mu $ on $Q$
by
\begin{equation}\label{337}
\langle  \alpha_\mu (q), v \rangle=  \langle \mu, \alpha (q, v) \rangle
\end{equation}
{\it i.e.,}
\begin{equation}\label{338}
(\alpha_\mu )_i  =  g_{ij} A^j_{\;b} \mu_a {\Bbb I}^{ab}.
\end{equation}
One sees from $\alpha(\xi _Q(q)) = \xi$ that $\alpha_\mu$ takes values in
${\bf J} ^{-1} (\mu)$.
The horizontal-vertical decomposition of a vector $(q, v) \in T_q Q$ is given
by
\begin{equation}\label{3310}
v  =  {\rm hor}_q v + {\rm ver}_q v
\end{equation}
where
\[ {\rm ver}_q v =  [\alpha (q, v)]_Q (q) \quad \mbox{and} \quad {\rm
hor}_q v =  v -
{\rm ver}_q v . \]
Notice that $ {\rm hor} : TQ \rightarrow {\bf J} ^{-1} (0)$ and as such, it
may be regarded as a
{\tenbi velocity shift\/}.

The {\tenbi amended potential\/} $V_\mu$ is defined by
\begin{equation}\label{3320}
V_\mu (q) = V (q) + \frac{1}{2} \langle  \mu, {\Bbb I} (q) ^{-1} \mu \rangle .
\end{equation}
In coordinates,
\begin{equation}\label{3319}
V_\mu (q)  =  V(q) + \frac{1}{2} {\Bbb I}^{ab} (q) \mu_a \mu_b.
\end{equation}

We recall from Abraham and Marsden [1978] or Simo, Lewis, and Marsden
[1991] that in a symplectic
manifold $ (P, \Omega )$,   a point $ z_e \in P $ is called a {\tenbi
relative equilibrium\/} if
\[ X_H(z_e) \in T_{z_e}(G \cdot z_e) \]
{\it i.e.,} if the Hamiltonian vector field at $ z_e $ points in the
direction of the group
orbit through $z_e$.
The {\it Relative Equilibrium Theorem\/} states that if $ z _e \in P $ and
$ z_e(t)$ is the
dynamic orbit of $ X_H $ with $ z_e(0) = z_e$ and $ \mu = {\bf J} (z _e) $,
then the following
conditions are equivalent
\begin{enumerate}
\item $ z_e $ is a relative equilibrium
\item $ z_e(t) \in G _\mu \cdot z _e \subset G \cdot z _e $
\item there is a $\xi \in \frak g $ such that $ z_e(t) = \exp (t \xi)\cdot z_e
$

\item there is a $\xi \in \frak g $ such that $ z_e $ is a critical point of
the
{\tenbi augmented Hamiltonian\/}
\begin{equation}\label{411}
H_\xi (z) := H(z) - \langle  {\bf J} (z) - \mu, \xi \rangle
\end{equation}
\item $ z_e $ is a critical point of $ H \times {\bf J} : P \rightarrow {\Bbb
R}
\times \frak g ^{\ast} $, the {\tenbi energy-momentum map\/}
\item $ z_e $ is a critical point of $ H | {\bf J} ^{-1} (\mu) $
\item $ z_e $ is a critical point of $ H | {\bf J} ^{-1} ({\cal O}) $, where $
{\cal O} = G \cdot \mu \in \frak g ^{\ast} $
\item $ [z_e] \in P _\mu $ is a critical point of the reduced Hamiltonian $ H
_\mu $.
\end{enumerate}

Straightforward algebraic manipulation shows that $ H _\xi $ can be rewritten
as
follows
\begin{equation}\label{421}
H _\xi (q, v) = K _\xi (q, v) + V _\xi (q) + \langle  \mu , \xi \rangle
\end{equation}
where
\begin{equation}\label{422}
K _\xi (q, v) = \frac{1}{2} \| v - \xi _Q (q) \| ^2 ,
\end{equation}
and where
\begin{equation}\label{423}
V _\xi (q) = V (q) - \frac{1}{2} \langle  \xi, {\Bbb I} (q) \xi \rangle.
\end{equation}

These identities show the following.

\begin{prop} A point $ z _e = (q _e, v _e) $ is a relative equilibrium if
and only if
there is a $ \xi \in \frak g $ such that
\begin{enumerate}
\item $ v_e = \xi _Q(q _e)$ and
\item $ q _e $ is a critical point of $ V _\xi $.
\end{enumerate}
\end{prop}

The functions $ K _\xi $ and $ V _\xi $ are called the {\tenbi augmented\/}
kinetic and
potential energies respectively. The main point of this proposition is that
{\it it reduces
the job of finding relative equilibria to finding critical points  of\/} $
V _\xi $.

Relative equilibria may also be characterized by the amended potential. One
has the following
identity:
\[
H (q, p) = K _\mu  (q, p) + V _\mu  (q)
\]
where
\[
K _\mu  (q, p) = \frac{1}{2} \| p - \alpha _\mu (q) \| ^2 ,
\]
for $ (q,p) \in {\bf J} ^{-1} (\mu )$. This leads to the following:

\begin{prop} A point $ (q _e, v _e)$ with $ {\bf J}(q _e, v
_e) = \mu $ is a relative equilibrium if and only if
\begin{enumerate}
\item $ v _e = \xi _Q  (q _e) $ where $ \xi = {\Bbb I} ^{-1} (q) \mu $  and
\item $ q _e $ is a critical point of $ V _\mu $.
\end{enumerate}
\end{prop}

Next, we summarize the energy-momentum method of Simo, Posbergh and Marsden
[1990, 1991], Simo,
Lewis and Marsden [1991], on the purely Lagrangian side, Lewis [1992], and
in a control
theoretic context, Wang and Krishnaprasad [1992]. This is a technique for
determining
the stability of relative equilibria and for putting the equations of
motion linearized at a
relative equilibrium, into normal form. This normal form is based on a
special decomposition into
rigid and internal variables.

We confine ourselves to the {\tenbi regular case\/}; that is, we
assume $ z _e $ is a relative equilibrium that is also a regular point
({\it i.e.,} $ \frak
g _{ z _e} = \{ 0\}$, or $ z _e $ has a {\it discrete\/} isotropy group)
and $ \mu
= {\bf J} (z _e) $ is a generic point in $ \frak g ^{\ast} $ ({\it i.e.,}
its orbit is of
maximal dimension). We are seeking conditions for stability of $ z _e $
modulo $ G _\mu $.

The {\tenbi energy-momentum method\/} is as follows:
Choose a subspace $ {\cal S} \subset {\rm ker} {\bf D} {\bf J} (q_e , v _e
) $ that is also
transverse to the $ G _\mu $ orbit of $ (q _e , v _e ) $
\\ \begin{tabular}{rl}
{\bf a} & find $ \xi \in \frak g $ such that $ \delta H _\xi (z _e) = 0 $ \\
{\bf b} & test $ \delta ^2 H _\xi (z _e) $ for definiteness on $ {\cal S}$.
\end{tabular}

\begin{thm} {\bf The Energy-Momentum Theorem}. If $ \delta ^2 H _\xi (z _e)
$ is definite,
then $ z _e $ is $ G _\mu $-orbitally stable in $ {\bf J} ^{-1} (\mu) $ and
$ G $-orbitally
stable in $P$.
\end{thm}

For simple mechanical systems, one way to choose $ {\cal S} $ is as follows.
Let
\[ {\cal V}  = \{ \delta q \in T _{ q _e} Q \mid \langle \! \langle  \delta
q, \chi _Q (q _e)
\rangle \! \rangle = 0 \quad \mbox{for all} \quad \chi \in \frak g _\mu \}, \]
the {\it metric\/} orthogonal complement of the tangent space to the $ G
_\mu $-orbit in $Q$. Let
\[ {\cal S} = \{ \delta z \in {\rm ker}\, {\bf D} {\bf J} (z _e) \mid T \pi
_Q \cdot \delta
z \in {\cal V}  \} \]
where $ \pi _Q: T ^\ast Q = P \rightarrow Q $ is the projection.

If the energy-momentum method is applied to mechanical systems with
Hamiltonian $ H $ of
the form kinetic energy $ (K)$ plus potential $ (V)$, under hypotheses
given below, it is
possible to choose variables in a way that makes the determination of
stability conditions
sharper and more computable. In this set of variables (with the
conservation of momentum
constraint and a gauge symmetry constraint imposed on $ {\cal S} $), the
second variation
of $ \delta ^2 H _\xi $ block diagonalizes; schematically

\[ \delta ^2 H _\xi =
\left[ \begin{array}{cc}
\left[ \begin{array}{c} {\rm coadjoint} \\ {\rm orbit \, block}
\end{array} \right]   & 0 \\
0 & \left[ \begin{array}{c} {\rm Internal \, vibration} \\ {\rm block}
\end{array} \right]   \end{array} \right] . \]

Furthermore, the internal vibrational block takes the form
\[ \left[ \begin{array}{c} {\rm Internal \, vibration} \\ {\rm block}
\end{array}
\right] = \left[ \begin{array}{cc} \delta ^2 V _\mu  & 0 \\ 0 & \delta ^2 K
_\mu  \end{array} \right]  \]
where $ V _\mu  $ is the amended potential defined earlier, and $ K _\mu $
is a momentum
shifted kinetic energy. Thus, formal stability is equivalent to $ \delta ^2
V _\mu > 0 $
and that the overall structure is stable when viewed as a rigid structure,
which, as far as
stability is concerned, separates out the overall rigid body motions from
the internal motions
of the system under consideration.

To define the rigid-internal splitting, we begin with a splitting in
configuration space.
Consider (at a relative equilibrium) the space $ {\cal V}  $ defined above
as the metric
orthogonal complement to $ \frak g _\mu \cdot q $ in $ T _q Q $. Here we
drop the subscript
$ e $ for notational convenience. Then we split
\begin{equation}\label{531}
{\cal V}  = {\cal V}  _{{\rm RIG}} \oplus {\cal V}  _{{\rm INT}}
\end{equation}
as follows. Define
\begin{equation}\label{532}
{\cal V}  _{{\rm RIG}} = \{ \eta _Q (q) \in T _q Q \mid \eta \in \frak g
^\perp_\mu \}
\end{equation}
where $ \frak g ^\perp_\mu $ is the orthogonal complement to $ \frak g _\mu
$ in $\frak g$
with respect to the locked inertia metric. (This choice of orthogonal
complement depends on
$q$, but we do not include this in the notation.) From~(\ref{531}) it is
clear that $ {\cal
V}  _{{\rm RIG}} \subset {\cal V}  $ and that $ {\cal V}  _{{\rm RIG}} $
has the dimension of
the coadjoint orbit through $\mu$. Next, define
\begin{equation}\label{533}
{\cal V}  _{{\rm INT}} = \{ \delta q \in {\cal V}  \mid \langle  \eta, [
{\bf D} {\Bbb I} (q)
\cdot \delta q ] \cdot \xi \rangle = 0 \quad \mbox{for all} \quad \eta \in
\frak g
^\perp_\mu  \} \end{equation}
where $ \xi = {\Bbb I} (q) ^{-1} \mu $. An equivalent definition is $ {\cal
V}  _{{\rm INT}}
= \{ \delta q \in {\cal V}  \mid [ {\bf D} {\Bbb I} (q) ^{-1} \cdot \delta
q] \cdot \mu \in
\frak g _\mu \} $. The definition of $ {\cal V}  _{{\rm INT}} $ has an
interesting mechanical
interpretation in terms of the objectivity of the centrifugal force in case
$ G = SO(3) $;
see Simo, Lewis and Marsden [1991].

Define the {\tenbi Arnold form\/} $ {\cal A} _\mu : \frak g ^\perp_\mu
\times \frak g
^\perp_\mu \rightarrow {\Bbb R} $ by
\begin{equation}\label{534}
{\cal A} _\mu (\eta, \zeta) = \langle  a d ^\ast _\eta \mu, \chi _{ (q,
\mu)}(\zeta)\rangle
= \langle  \mu, a d _\eta \chi _{(q, \mu)} (\zeta) \rangle ,
\end{equation}
where $ \chi _{(q, \mu) }: \frak g ^\perp_\mu \rightarrow \frak g $ is
defined by
$ \chi _{(q, \mu)} (\zeta) = {\Bbb I} (q) ^{-1} a d ^\ast _\zeta \mu + a d
_\zeta {\Bbb I} (q)
^{-1} \mu $. The Arnold form appears in Arnold's [1966] stability analysis of
relative
equilibria in the special case $ Q = G $. At a relative equilibrium, the
form $ {\cal A}
_\mu $ is symmetric, as is verified either directly or by recognizing it as
the second
variation of $ V _\mu $ on $ {\cal V}  _{{\rm RIG}} \times {\cal V}  _{{\rm
RIG}}$.

At a relative equilibrium, the form $ {\cal A} _\mu $ is degenerate as a
symmetric bilinear
form on $ \frak g ^\perp_\mu $ when there is a non-zero $ \zeta \in \frak g
^\perp_\mu $
such that
\[ {\Bbb I} (q) ^{-1} a d ^\ast _\zeta \mu + a d _\zeta {\Bbb I} (q)
^{-1} \mu \in \frak g _\mu ; \]
in other words, when $ {\Bbb I} (q)^{-1}: \frak g ^{\ast} \rightarrow \frak
g $ has a
{\it nontrivial symmetry\/} relative to the (coadjoint, adjoint) action of
$\frak g$
(restricted to $ \frak g ^\perp_\mu $) on the space of linear maps from
$\frak g ^{\ast}$
to $\frak g$. (When one is not at a relative equilibrium, we say the Arnold
form is {\it
non-degenerate \/} when $ {\cal A} _\mu (\eta, \zeta) = 0 $ for all $ \eta
\in \frak g
^\perp_\mu $ implies  $ \zeta  = 0$.) This means, for $ G = SO(3)$ that $
{\cal A} _\mu $
is non-degenerate if $\mu$ is not in a multidimensional eigenspace of $
{\Bbb I} ^{-1} $.
Thus, {\it if the locked body is not symmetric ({\it i.e.,} a Lagrange
top), then the
Arnold form is non-degenerate\/}.

\begin{prop} If the Arnold form is non-degenerate, then
\begin{equation}\label{535}
{\cal V}  = {\cal V}  _{{\rm RIG}} \oplus {\cal V}  _{{\rm INT}}.
\end{equation}
\end{prop}
Indeed, non-degeneracy of the Arnold form implies $ {\cal V} _{{\rm RIG}}
\cap {\cal V}_{{\rm
INT}} = \{ 0\}$ and, at least in the finite dimensional case, a dimension count
gives~(\ref{535}). In the infinite dimensional case, the relevant
ellipticity conditions are
needed.

The split~(\ref{535}) can now be used to induce a split of the phase space
\begin{equation}\label{536}
{\cal S} = {\cal S}_{{\rm RIG}} \oplus {\cal S}_{{\rm INT}}.
\end{equation}
Using a more mechanical viewpoint, Simo, Lewis and Marsden [1991] show how
$ {\cal S} _{{\rm
RIG}}$ can be defined by extending $ {\cal V}  _{{\rm RIG}} $ from
positions to momenta using
{\it superposed rigid motions\/}. For our purposes, the important
characterization of $
{\cal S} _{{\rm RIG}} $ is via the mechanical connection:
\begin{equation}\label{537}
{\cal S} _{\rm RIG} = T _q \alpha _\mu \cdot {\cal V}  _{\rm RIG}
\end{equation}
so $ {\cal S} _{\rm RIG} $ is isomorphic to $ {\cal V}  _{\rm RIG} $. Since
$ \alpha _\mu $
maps $Q$ to $ {\bf J} ^{-1} (\mu) $ and $ {\cal V}  _{\rm RIG} \subset
{\cal V}  $, we get $
{\cal S} _{\rm RIG} \subset {\cal S} $. Define
\begin{equation}\label{538}
{\cal S} _{\rm INT} = \{ \delta z \in {\cal S} \mid \delta q \in {\cal V}
_{\rm INT} \};
\end{equation}
then~(\ref{536}) holds if the Arnold form is non-degenerate. Next, we write
\begin{equation}\label{539}
{\cal S} _{\rm INT} = {\cal W} _{\rm INT} \oplus {\cal W} ^{\dagger} _{\rm
INT} ,
\end{equation}
where $ {\cal W} _{\rm INT} $ and $ {\cal W} ^\dagger _{\rm INT} $ are
defined as follows:
\begin{equation}\label{5310}
{\cal W} _{\rm INT} = T _q \alpha _\mu \cdot {\cal V}  _{\rm INT} \quad
\mbox{and} \quad
{\cal W} ^\dagger _{\rm INT} = \{ {\rm ver} (\gamma) \mid \gamma \in [\frak
g \cdot q]^0 \}
\end{equation}
where $ \frak g \cdot q = \{ \zeta _Q (q) \mid \zeta \in \frak g \}, \quad
[ \frak g \cdot
q]^0 \subset T ^\ast _q Q $ is its annihilator, and $ {\rm ver}(\gamma) \in
T _z (T ^\ast Q) $
is the vertical lift of $ \gamma \in T ^\ast _q  Q $; in coordinates, ${\rm
ver}(q ^i,
\gamma _j) = (q ^i, p _j, 0, \gamma _j) $. The vertical lift is given
intrinsically by
taking the tangent to the curve $ \sigma (s) = z + s \gamma $ at $ s = 0 $.

\begin{thm} {\bf Block Diagonalization Theorem} Assume that the Arnold form is
nondegenerate. Then in the splittings introduced above at a relative
equilibrium, $ \delta
^2 H _\xi (z _e) $ and the symplectic form $ \Omega _{ z _e} $ have the
following form:
\begin{eqnarray*}
 \delta ^2 H _\xi (z _e) = \left[ \begin{array}{ccc} \left[ \begin{array}{c}
{\rm Arnold} \\ {\rm form} \end{array}\right] & 0 & 0 \\ 0 & \delta ^2 V
_\mu  & 0 \\ 0 & 0 & \delta ^2 K _\mu \end{array} \right]
\end{eqnarray*}
and
\begin{eqnarray*} \quad \Omega _{ z _e } = \left[ \begin{array}{ccc} \left[
\begin{array}{c} {\rm coadjoint \: orbit} \\ {\rm symplectic \: form}
\end{array}\right] & \left[ \begin{array}{c}
{\rm internal \: rigid} \\ {\rm coupling} \end{array}\right] & 0 \\ & & \\
- \left[
\begin{array}{c} {\rm internal\:rigid} \\ {\rm coupling} \end{array}\right] & S
& {\cal I}  \\ & & \\ 0 & - {\cal I}^T   & 0 \end{array} \right]
\end{eqnarray*}
where the columns represent elements of ${\cal S}_{\rm RIG}, {\cal W} _{\rm
INT}$  and
$ {\cal W} _{\rm INT} ^\dagger $, respectively, and $ {\cal I} : {\cal W}
_{\rm INT} ^\dagger
\rightarrow {\cal W} _{\rm INT} ^{\ast} $ is the isomorphism given as
follows: Let $ {\rm
vert}(\gamma ) \in {\cal W} _{\rm INT} ^\dagger $ where $ \gamma \in [\frak
g \cdot q] ^0 $ and
let $ \delta q \in {\cal V} _{\rm INT} $; then
\[
\left\langle {\cal I}({\rm vert} (\gamma)), T _q \alpha _\mu \cdot \delta q
\right\rangle =
\left\langle \gamma, \delta q \right\rangle.
\]
\end{thm}

As far as stability is concerned, we have the following consequence of block
diagonalization.

\begin{thm} {\bf Reduced Energy-Momentum Method} Let $ z _e = (q _e, p _e)
$ be a
(cotangent) relative equilibrium and assume that the internal variables are
not trivial;
{\it i.e.,} $ {\cal V} _{\rm INT} \neq \{ 0\} $. If $ \delta ^2 H _\xi (z
_e) $ is
definite, then it must be positive definite. Necessary and sufficient
conditions for $
\delta ^2 H _\xi (z _e) $ to be positive definite are
\begin{enumerate}
\item the Arnold form is positive definite on $ {\cal V} _{\rm RIG} $ and
\item $ \delta ^2 V _\mu (q _e) $ is positive definite on $ {\cal V} _{\rm
INT} $.
\end{enumerate}
\end{thm}

This follows since $ \delta ^2 K _\mu $ is positive definite and $ \delta ^2 H
_\xi $ has the above block diagonal structure.

In examples, it is this form of the energy-momentum method that is normally
easiest to
use.

A straightforward calculation establishes the useful relation
\begin{equation}\label{5316}
\delta ^2 V _\mu (q _e) \cdot (\delta q, \delta q) = \delta ^2 V _\xi (q
_e) \cdot
(\delta q, \delta q) + \langle ({\bf D} {\Bbb I} (q _e) \cdot \delta q)
\xi, ({\Bbb I} (q
_e) ^{-1} \circ  {\bf D} {\Bbb I} (q _e) \cdot \delta q) \xi  \rangle
\end{equation}
and the correction term is positive. Thus, if $ \delta ^2 V _\xi (q _e) $
is positive
definite, then so is $ \delta ^2 V _\mu (q _e) $, but not necessarily
conversely. {\it
Thus, $ \delta ^2 V _\mu (q _e)$ gives sharp conditions for stability (in
the sense of
Theorem {\bf 2.3}), while $ \delta ^2 V _\xi $ gives only sufficient
conditions.\/}

Using the notation $ \zeta = [ {\bf D}  {\Bbb I} ^{-1} (q _e) \cdot \delta
q] \mu \in
\frak g _\mu $ (see the comments following~(\ref{533})), observe that the
``correcting
term'' in~(\ref{5316}) is given by
$ \langle  {\Bbb I} (q _e) \zeta, \zeta \rangle = \langle \! \langle
\zeta _Q (q _e), \zeta _Q (q _e) \rangle \! \rangle $.

One of the most interesting aspects of block diagonalization is that the
rigid-internal
splitting also brings the symplectic structure into
normal form. We already gave the general structure of this and here we
provide a few
more details. We emphasise once more that this implies that the equations
of motion are
also put into normal form and this is useful for studying eigenvalue
movement for
purposes of bifurcation theory. For example, for {\it abelian\/} groups,
the linearized
equations of motion take the {\it gyroscopic form\/}:
\[ M  \ddot{q} + S \dot{ q} + \Lambda q = 0 \]
where $ M $ is a positive definite symmetric matrix (the mass matrix), $
\Lambda $ is
symmetric (the potential term) and $ S $ is skew (the gyroscopic, or
magnetic term).
This second order form is particularly useful for finding eigenvalues of
the linearized
equations (see, for example, Bloch, Krishnaprasad, Marsden and Ratiu [1991]).

To make the normal form of the symplectic structure explicit, we need some
preliminary results.
See Simo, Lewis and Marsden [1991] for the proofs.

\begin{lem} Let $ \Delta q = \eta _Q (q _e) \in {\cal V} _{\rm RIG} $ and $
\Delta z =
T \alpha _\mu \cdot \Delta q \in {\cal S} _{\rm RIG} $. Then
\begin{equation}\label{541}
\Delta z = {\rm vert} \, [ {\Bbb F} L (\zeta _Q (q _e))] - T ^\ast \eta _Q
(q _e) \cdot
p _e
\end{equation}
where $ \zeta = {\Bbb I} (q _e) ^{-1} a d ^\ast _\eta \mu $, {\rm vert}
denotes the
vertical lift, and $ {\Bbb F}L$  denotes the fiber derivative.
\end{lem}

\begin{lem} For any $ \delta z \in T _{z _e } P $,
\begin{equation}\label{549}
\Omega (z _e) (\Delta z, \delta z) = \langle  [ {\bf D} {\bf J} (z _e)
\cdot \delta z
] , \eta \rangle - \langle \! \langle \zeta _Q (q _e), \delta q \rangle \!
\rangle .
\end{equation}
\end{lem}

If $ \delta z \in {\cal S}_{\rm INT} $, then it lies in ker ${\bf D} {\bf
J} $, so we get
the {\tenbi internal-rigid interaction terms\/}:
\begin{equation}\label{5410}
\Omega (z _e) (\Delta z, \delta z) = - \langle \!
\langle  \zeta _Q (q _e), \delta q \rangle \! \rangle = - \left\langle \mu
_e , [\eta , \alpha(\delta
q )] \right\rangle .
\end{equation}
Since these involve only $ \delta q $ and not $ \delta p $, there is a zero
in the last
slot in the first row of $\Omega $ and so we can define the operator $C$
by (2.34): $ \left\langle C(\delta q), \Delta q \right\rangle := \Omega (z
_e) (\Delta z, \delta
z)$.

\begin{lem} The {\tenbi rigid-rigid terms\/}
in $ \Omega $ are
\begin{equation}\label{5411}
\Omega (z _e) (\Delta _1 z, \Delta _2 z) = - \langle  \mu, [\eta _1, \eta
_2] \rangle ,
\end{equation}
which is the coadjoint orbit symplectic structure.
\end{lem}

Next, we turn to the magnetic terms:

\begin{lem} Let $ \delta _1 z = T \alpha _\mu \cdot \delta _1 q $ and $
\delta _2 z = T
\alpha _\mu \cdot \delta _2 q \in {\cal W} _{\rm INT} $, where $ \delta _1
q, \delta _2 q
\in {\cal V} _{\rm INT} $. Then
\begin{equation}\label{5412}
\Omega (z _e) (\delta _1 z, \delta _2 z) = - {\bf d} \alpha _\mu (\delta _1
q, \delta _2 q)
\end{equation}
\end{lem}

If we define the one form $ \alpha _\xi $ by $ \alpha _\xi (q) = {\Bbb F} L
(\xi _Q (q)) $,
then the definition of $ {\cal V} _{\rm INT} $ shows that on this space $
{\bf d} \alpha
_\mu = {\bf d} \alpha _\xi $. This is a useful remark since $ {\bf d}
\alpha _\xi $ is
somewhat easier to compute in examples. We also note, as in an earlier
remark, that the magnetic
terms can be equivalently computed from the magnetic terms of the $ G _\mu
$ connection
rather that the $G$ connection. For instance, for the water molecule, this
is easier since
in that case, $G = SO(3) $ while $ G _\mu = S ^1 $.
\vspace{0.2in}

Let us now introduce a change of variables $r \mapsto  T \pi _Q \cdot r$ of
${\cal
S} _{\rm RIG} $ to $ {\cal V} _{\rm RIG} $ and $ p \mapsto {\cal I} p $ of
${\cal W}
_{\rm INT} ^{\dagger} $ to $ {\cal W} ^{\ast} _{\rm INT} $, so that the
representations for $\delta ^2 H _\xi (z _e)$ and $\Omega _{ z _e}$ will be
relative
to the space ${\cal V} _{\rm RIG} \oplus {\cal W} _{\rm INT} \oplus {\cal
W} _{\rm
INT} ^{\ast}$. We note at this point that we could have equally well used the
representation relative to ${\cal V} _{\rm RIG} \oplus {\cal V} _{\rm INT}
\oplus
{\cal V} _{\rm INT} ^{\ast}$ and the results below would not materially change
(replace ${\cal W}$ by ${\cal V}$ where appropriate). Using this
representation,
introduce the following notation for the block diagonal form of $ \delta ^2
H_{\xi}:
$  \begin{equation}  \left[ \begin{array}{ccc} A_{\mu } & 0 & 0 \\ 0 &
\Lambda  & 0 \\
0 & 0 & M^{-1} \end{array} \right] \end{equation}  where $A _\mu $ is the
co-adjoint
orbit block;  {\it i.e.,\/} the Arnold form, $ (2 \times 2
$ in the case of $ G = SO(3)), \Lambda $
corresponds to the second variation of the amended
potential energy, and $M$ corresponds to the metric on the internal variables.

The corresponding symplectic form for the linearized dynamics is
\begin{equation}
\Omega = \left[ \begin{array}{ccc} L_{\mu } & C & 0 \\ -C^T & S  & {\bf 1} \\ 0
&
-{\bf 1} & 0 \end{array} \right],
\end{equation}
where $S$ is skew-symmetric, $ {\bf 1} $ is the identity and
where $ C : {\cal W} _{\rm INT} \rightarrow {\cal V} ^{\ast} _{\rm RIG} $
is defined by (2.34).
 From the earlier remarks, note that in (2.37) and (2.38), the upper block
corresponds to the
``rotational'' dynamics ($L_{\mu}$ is in fact the co-adjoint orbit
symplectic form for $G$) while
the two lower blocks correspond to the ``internal'' dynamics.  In (2.38)
$C$ represents coupling
between the internal and rotational dynamics, while $S$ gives the Coriolis
or gyroscopic forces.

The corresponding linearized Hamiltonian vector field is then given by
\[ X_H =( \Omega ^{-1})^T \nabla  H = (\Omega ^{-1})^T \delta ^2 H_{\xi}, \]
which a computation given below reveals to be
\begin{equation}
X_H(r,q,p) = \left[ \begin{array}{ccc} -L_{\mu }^{-1} A _{\mu} & 0 &
-L_{\mu}^{-1} C M ^{-1} \\
{}\\
 0 & 0 &  M^{-1} \\ {}\\
 - C^T L^{-1}_{\mu} A_{\mu} & - \Lambda  & - \tilde{S} M^{-1} \end{array}
\right]
\left[ \begin{array}{c} r \\{}\\ q \\{} \\ p  \end{array} \right]
\end{equation}
where $ \tilde{S} = S + C^T L_{\mu }^{-1} C = -\tilde{S}^T. $

Thus our linearized equations have the form
\begin{equation}
\dot{z} = X_H(z)
\end{equation}
where $ z = (r, q, p) \in {\cal V} _{\rm RIG} \times {\cal W} _{\rm INT}
\times {\cal W} _{\rm
INT} ^{\ast} $ and $ X_H $ is given by (2.39). See Lewis [1993] and Lewis
and Ratiu [1993] for
the explicit expression (in terms of the basic data) of the linearized
equations. To
prove (2.39) we use the first part of the following lemma.  (The remainder
of the lemma will be
used in our stability calculations.)

\begin{lem} Consider a vector space $ V = V _1 \oplus V _2 $ and a linear
operator $ M : V
\rightarrow V ^{\ast} $ given by the partitioned matrix
\[
 M = \left[ \begin{array}{cc} B _{11} & B
_{ 12} \\ B _{ 21} & B _{ 22}  \end{array} \right]
\]
where $B _{ 11} : V _1 \rightarrow V _1 ^{\ast}, B _{ 12} : V _2
\rightarrow V _1 ^{\ast},  B _{
21} : V _1 \rightarrow V _2 ^{\ast} $ and $ B _{ 22} : V _2 \rightarrow V
_2 ^{\ast} $ are linear
maps.  Assume $B _{ 11}$  is an isomorphism and let
\[
 N = \left[ \begin{array}{cc} B _{ 11} & 0  \\
0 & B _{ 22} - B _{ 21}B _{ 11}^{-1}B _{ 12}
 \end{array} \right],  L = \left[ \begin{array}{cc} {\bf 1}  & 0  \\ -B _{
21}B _{ 11}^{-1} &
{\bf 1}  \end{array} \right], P =   \left[ \begin{array}{cc}
{\bf 1} & -B _{ 11}^{-1}B _{ 12}  \\ 0 & {\bf 1}  \end{array} \right],
\]
so that $ N : V \rightarrow V ^{\ast}, L : V ^{\ast} \rightarrow V ^{\ast}$
and $ P : V \rightarrow
V $. Then \begin{description}
\item [\ \ (i)] $ L M P = N $
\item [\ (ii)]  $M$ is symmetric if and only if $B _{ 11}$ and $B _{ 22}$
are symmetric
and $ B _{ 12}^T = B _{ 21} $
\item [(iii)]  If $M$  is symmetric so is $N$
\item [(iv)]  If $M$  is symmetric then it is positive definite iff $N$ is
positive definite; more generally, the signatures of $M$ and $N$ coincide.
\end{description}
\end{lem}

\noindent{\bf Proof} \quad {\bf (i)} is a computation, while {\bf (ii)} and
{\bf
(iii)} are obvious.  For {\bf (iv)}, note first that $ L^T = P. $ If $
\langle \,, \rangle  $ denotes the natural pairing, then
\[
\langle L M P x, x \rangle = \langle M P x, L^ T x \rangle =
\langle M P x, P x \rangle,
\]
which shows that $N$ and $M$ have the same signature since $P$ is
invertible. \quad $\blacksquare$

\begin{lem} The linearized Hamiltonian flow with Hamiltonian $ \delta^2 H_{\xi}
$ is given by  $(2.39)$.  \end{lem}

\noindent {\bf Proof} \quad We have  $ X_{\delta^{2} H_{\xi}} = ( \Omega
^{-1}) ^T \delta^2 H_{\xi}. $   To invert $ \Omega $, set $
B _{ 11} = L_{ \mu },\,\,
B _{ 12} = [C \; \; 0],\,
B _{ 21} = \left[ \begin{array}{c} -C^T \\ 0 \end{array} \right],$  and
$ B _{ 22} = \left[ \begin{array}{cc}
S & {\bf 1}  \\ -{\bf 1}  & 0   \end{array} \right]. $
 From part {\bf (i)} of Lemma {\bf 2.11}, we have $ M^{-1} = V N^{-1} L$
which gives
\[ \Omega ^{-1} =  \left[ \begin{array}{ccc} {\bf 1}  & -L_{\mu} ^{-1} C & 0 \\
0 &   {\bf 1} & 0  \\0 &  0 & {\bf 1} \end{array} \right]
\left[ \begin{array}{ccc}
L_{\mu} & 0 & 0 \\ 0 &  \tilde{S} & {\bf 1}  \\ 0 & - {\bf 1}  & 0
\end{array} \right] ^{-1}
 \left[ \begin{array}{ccc} {\bf 1} & 0 & 0 \\ C^TL_{\mu
}^{-1} &   {\bf 1} & 0  \\ 0 & 0 & {\bf 1}  \end{array} \right] \]
where $ \tilde{ S} = S + C ^T L _\mu ^{-1} C = - \tilde{ S} ^T $. Noting that
\[ \left[ \begin{array}{cc} \tilde{ S} & {\bf 1}  \\ -{\bf 1} & 0
\end{array} \right]^{-1}  = \left[ \begin{array}{cc} 0 & - {\bf 1}  \\ {\bf 1}
& \tilde{ S} \end{array} \right] \]
we obtain
\[ \Omega ^{-1} = \left[ \begin{array}{ccc} L_{\mu }^{-1} & 0 & L_{\mu}^{-1}C
\\
0 & 0 & - {\bf 1}  \\  C^T L_{\mu}^{-1} & {\bf 1}   &  \tilde{S}  \end{array}
\right] \]
and
\[ ( \Omega ^{-1}) ^T = \left[ \begin{array}{ccc} -L_{\mu}^{-1} & 0 &
-L_{\mu}^{-1}C \\ 0 & 0 & {\bf 1}  \\ -   C^T
L_{\mu}^{-1} & - {\bf 1}   & - \tilde{S}
\end{array} \right]
\]
Hence we get result. \quad $\blacksquare$

\section{The Chetaev Function and Lyapunov Instability}

In this section we add a (small) dissipation term to the linear Hamiltonian
equation (2.40) and show that this results in Lyapunov instability for the
linear system.  This is insufficient to prove nonlinear instability of the
original system about the given relative equilibrium.  For this we
prove a result on spectral instability, which we do in \S {\bf 4}.

We add dissipation (damping) to the ``internal'' variables of the system
only, in accordance with the natural physical models.  The dissipation is
assumed to occur due to the addition, to the Lagrangian, of a Rayleigh
dissipation function (see
e.g. Whittaker [1959]):
\begin{equation}
{\cal R} = \frac{1}{2} \dot{q}^T R\dot{q} = \frac{1}{2}( M^{-1} p) ^T  R
M^{-1} p,
\end{equation}
where the Rayleigh dissipation matrix $ R : {\cal W} _{\rm INT} \rightarrow
{\cal W} _{\rm INT}
^{\ast} $ is symmetric and positive definite: $ R = R^T \geq 0$.

The system of linearized equations (2.40) becomes

\begin{equation} \left. \begin{array}{ccl}
\dot{r} & =   & - L_{\mu} ^{-1} A_{\mu} r -L_{\mu} ^{-1}C M ^{-1} p  \\
\mbox{ } \\
\dot{q} & = & M^{-1} p  \\
\mbox{ } \\
 \dot{p} & = & \left( -C^TL_{\mu}  ^{-1} A_{\mu} r - \Lambda q - \tilde{S}
M^{-1} p  - R M^{-1} p \right).  \end{array} \right\}\end{equation}
We note that
\begin{equation}
\frac{ d \delta^2 H_{\xi}}{dt}  = - 2 {\cal R}.
\end{equation}

The presence of dissipation results in the addition of a term $ - R M^{-1}
$ to the $ (3, 3)$
block of the matrix representation (2.39) of the linear system (2.40).

To prove Lyapunov instability, we will employ a generalization of the Chetaev
function (Chetaev [1961]; see also Arnold [1987]).

\vspace{0.2in}

Before doing the general case, it is instructive to analyze the special case $
G = S^1, $ an abelian group, where our system reduces to the form of the
system originally analyzed by Chetaev and Thomson. This analysis is relevant,
for example, for examining planar rotating systems (see e.g. Oh et al. [1989]).

In this case the $A_\mu $ block in (2.37) vanishes and, the
linearized flow $ X _{\delta ^2 H_{\xi}} $ with the addition of (internal)
damping becomes

\begin{equation} \left. \begin{array}{cc}
\dot{q} = M ^{-1} p \\
\mbox{ } \\
\dot{p} =   (- \Lambda q - (S + R)M^{-1} p) \end{array} \right\}
\end{equation}
where $ R = R^T \geq 0 $ is the Rayleigh dissipation matrix as above and $ S =
- S^T $ represents the gyroscopic forces in the system.

We shall call (3.4) the {\tenbi Chetaev-Thomson normal form\/}.  The
example (1.9)
analyzed in the introduction is the simplest case of this form.

The basic question addressed by Chetaev is the following.  If $ \Lambda $ has
some negative eigenvalues, yet the spectrum of
\begin{eqnarray*}
\dot{q} & = & M ^{-1} p  \\
\dot{p} & = &  - \Lambda q - S M ^{-1} p  \end{eqnarray*}
is on the imaginary axis, is the system (3.4) unstable?  Chetaev showed that
this is indeed the case for {\tenbi strong damping\/}; that is, when $ R $ is
positive definite.  Our proof is a slight modification of his.  Interestingly,
no assumption on $ S $ or the the size of $ R $ is explicitly needed.

\begin{thm} Suppose $ \Lambda $ is has one or more negative
eigenvalues and $ R $ is positive definite.  Then the system (3.4) is
(Lyapunov)
unstable. \end{thm}

The proof is based on the following.

\begin{lem} (Lyapunov's Instability Theorem)  A linear system is Lyapunov
unstable if there is a quadratic function $W$ whose associated quadratic
form has at
least one negative eigendirection and is such that $\dot{W}$ is negative
definite.\end{lem}

See, for example, LaSalle and Lefschetz [1963] for the proof of this lemma.
\vspace{0.2in}

To utilize this lemma to prove the theorem, we first assume that $\Lambda$ is
an isomorphism. Let
\begin{equation} W(q, p) = H_0(q, p) + \beta B q \cdot M ^{-1} p,
\end{equation}
where $ H_0 $ is the Hamiltonian for the undamped system, $ H_0 = \frac{1}{2}
p ^T M ^{-1} p + \frac{1}{2} q^T  \Lambda q$, $ \beta $
is a scalar, and $B$ is a linear map, both of which are to be determined.
Write

\begin{equation}
W = \frac{1}{2} \left( p^T\, q^T\right) \left[ \begin{array}{cc} M ^{-1}
& \beta (M ^T ) ^{-1} B \\ \beta  B^T M ^{-1} &
\Lambda \end{array} \right] \left( \begin{array}{c}  p\\ q\end{array}
\right) . \end{equation}
Calculating the time derivative, we find that
\begin{eqnarray}
\lefteqn{\dot{W} = - (p^T\,q^T) \times } \\
& &  \left[ \begin{array}{cc}
(M ^T )^{-1} RM^{-1}
&  \displaystyle{\frac{\beta}{2}}(M ^T ) ^{-1}(R -  S)M ^{-1} B \\
 - \displaystyle{\frac{\beta}{2}}((M ^T )^{-1}B M^{-1} + (M^T)^{-1} B^T M
^{-1})
& \\ \\
\\ \displaystyle{\frac{\beta } {2}}B ^T (M ^T ) ^{-1}(R + S)M ^{-1}
&  \displaystyle{\frac{\beta } {2}}(B ^T  M^{-1} \Lambda + \Lambda
(M ^T )^{-1} B) \end{array} \right] \left( \begin{array}{c} p\\
\\ \\ q\end{array} \right). \nonumber
\end{eqnarray}

Now choose any positive definite symmetric map $ K : {\cal W} _{\rm INT}
\rightarrow {\cal
W} _{\rm INT} ^{\ast} $. Our choice of $B$ will depend on $K$, but $K$ may
be chosen
arbitrarily, and this freedom will be important below. We let
\[
B = M ^T K^{-1} \Lambda :{\cal W}_{\rm INT}\rightarrow{\cal W}_{\rm
INT}^{\ast} .
\]
 From Lemma {\bf 2.11}, we see that $ \dot{ W} $ is negative definite if and
only if
\[
\left[ \begin{array}{cc} (M ^T ) ^{-1}R M ^{-1}+ O(\beta)
& 0 \\
0 & \beta \Lambda K ^{-1} \Lambda + O(\beta ^2)  \end{array} \right]
 \]
is positive definite. This is clearly true for $ \beta > 0 $ sufficiently
small since
$ (M^T ) ^{-1} R M ^{-1} $ and $ \Lambda K ^{-1} \Lambda $ are symmetric and
postive definite.  On the other hand, by a
similar argument, $ W $ has at least one negative eigendirection for $
\beta $ sufficiently
small.  Hence by lemma 3.2, we have instability.

To prove the general case, in which $\Lambda$ is allowed to be degenerate,
we proceed
as follows. Split the space
\[
{\cal W} _{\rm INT} = {\rm ker}\Lambda \oplus ({\rm ker}\Lambda) ^{\perp}
\]
into the direct sum of the kernel of $\Lambda$ and its orthogonal complement
in the inner product corresponding to $M$.
This induces a similar decompostion of the
dual spaces using $M$ as an isomorphism. Denote with a subscript $ 1 $ the
first
component in this decomposition and with a subscript $ 2 $, the second
component. In
this decomposition, we have the block structure
\[
\Lambda = \left[ \begin{array}{cc} 0 & 0 \\ 0 & \Lambda _2  \end{array} \right]
\]
and
\[
M = \left[ \begin{array}{cc} M _1  & 0 \\ 0 & M _2  \end{array} \right] .
\]
The equations (3.4) in this splitting become
\begin{eqnarray*}
 \left. \begin{array}{cc}
\dot{q _1 } = M_1  ^{-1} p _1  \\
\mbox{ } \\
\dot{q _2 } = M_2  ^{-1} p _2  \\
\mbox{ } \\
\dot{p} =   ((0,- \Lambda _2  q_2)  - (S + R)M^{-1} p). \end{array} \right\}
\end{eqnarray*}
Notice that the first equation for $ q _1 $ decouples from the next three
equations.
Now we proceed as above, with the function $ W (q,p) $ replaced by the
following function of $ (q _2, p) $:
\[
W(q _2, p) = \frac{1}{2} p ^T M ^{-1} p + \frac{1}{2} q _2 ^T \Lambda q _2
+ \beta M ^T _2 K _2 ^{-1} \Lambda _2 q _2 \cdot (M ^T  _2) ^{-1} p _2,
\]
where $ K _2 : ({\rm ker}\Lambda) ^{\perp} \rightarrow M({\rm ker}\Lambda)
^{\perp} )
\subset {\cal W} _{\rm INT} ^{\ast} $ is positive definite symmetric. Note
especially
that here we are using our freedom to choose $K$; in Chetaev, the initial
choice $ K =
\Lambda $ was made, which required $\Lambda$ to be invertible.  We now
compute $ \dot{
W}$ as above, and obtain an expression similar to (3.7) but with the blocks
done
according to the variables $ (p, q _2)$, and in which the top right and
lower left
expressions are modified, but are still multiplied by $\beta$, and where
the lower
right hand block is replaced by the expression  $ \beta \Lambda _2 K _2
^{-1} \Lambda
_2 $. Now repeat the argument above. \quad $\blacksquare$

\vspace{0.2in}

We now extend our analysis to the general equation (2.40) {\it i.e.,\/}
to an arbitrary nonabelian symmetry group $G.$  We show that
indefiniteness of $ \delta ^2 H_{\xi} $ at a given relative equilibrium implies
Lyapunov instability (again spectral instability follows from the analysis in
\S {\bf 4}).

The main ingredient is a generalization of the Chetaev function (3.5).
As above, we establish definiteness of the time derivative of
the function, but the analysis is now more complex.  Also we need an
assumption on the coupling matrix $C$  between the internal and rotational
modes.

\begin{thm}   Suppose $ A_{\mu} $ is nondegenerate and either $\Lambda$ or
$A _\mu$
has at least one negative eigenvalue.  Suppose that $ R > 0 $ and that $C^T$ is
injective. Then the system {\rm (3.2)} is Lyapunov unstable. \end{thm}

\noindent{\bf Proof} \quad As in the abelian case, we start with the
assumption that
$\Lambda$ is an isomorphism. In this case, let
\begin{eqnarray}  W(q, p, r) & = &
\frac{1}{2}p \cdot M^{-1}p +  \frac{1}{2} q \cdot \Lambda
q + \frac{1}{2} r\cdot  A_{\mu}r \nonumber \\
&  & + \,\beta B q \cdot M^{-1} p + \alpha D r
\cdot M^{-1} p + \gamma  E r \cdot \Lambda q \end{eqnarray}
where $ \alpha, \beta $, and $ \gamma $ are scalars and $ B, D  $, and $ E
$ are linear
operators, all to be chosen. We write the matrix representation of $ W, $ in
the
ordering $ (p, q, r) $ as :
\begin{equation}
W = \frac{1}{2} \left[ \begin{array}{ccc} M^{-1} & \displaystyle{\frac{\beta }
{2}} (M ^T )^{-1} B & \displaystyle{\frac{\alpha  } {2}} (M ^T )^{-1}D \\
\\
\displaystyle{\frac{\beta } {2}} B ^T M^{-1} & \Lambda  &
\displaystyle{\frac{\gamma  } {2}} \Lambda E \\
\\
 \displaystyle{\frac{\alpha  }
{2}} D^TM^{-1} & \displaystyle{\frac{\gamma  } {2}} E^T \Lambda  & A_{\mu }
\end{array} \right].
\end{equation}
After a lengthy computation, we find that the matrix representation of
$-\dot{W}$ is:
\begin{equation}
\hspace{-.65in}\left[
\begin{array}{ccc} %Big array
\begin{array}{c} %11 array
(M ^T ) ^{-1} R M ^{-1}  \\
-\frac{ \beta }{2} \left( (M^T )^{-1}B^T M ^{-1} \right. \\
+ \left. (M ^T ) ^{-1} B M^{-1}\right) \\
+ \frac{\alpha }{2} \left( - (M^T )^{-1} C^T L_{\mu }^{-1} D^T M ^{-1}
\right. \\
\left. + \, (M ^T ) ^{-1} D L_{\mu }^{-1} C M^{-1} \right)   \end{array} &
\begin{array}{c} %12 array
 \frac{\beta}{2}(M ^T ) ^{-1} (R - \tilde{ S} ) M ^{-1} B  \\
 - \frac{\gamma}{2} (M^T)^{-1} C^T L_{ \mu }^{-1} E^T \Lambda
\end{array} &
\begin{array}{c} %13 array
\frac{\alpha}{2}(M^T )^{-1} (R-\tilde{ S})M^{-1}D\\
 + \frac{\alpha}{2} M^{-1} D L_{\mu} ^{-1} A_{\mu} \\
 - \frac{\gamma }{2} (M^T )^{-1} \Lambda E
\end{array} \\
\mbox{  }  \\
\begin{array}{c}  %21 array
\left( \frac{\beta}{2}(M ^T ) ^{-1} (R - \tilde{ S} ) M ^{-1} B \right. \\
\left. - \frac{\gamma}{2} (M^T)^{-1} C^T L_{ \mu }^{-1} E^T \Lambda \right) ^T
\end{array} &
\begin{array}{c} %22 array
\frac{\beta}{2} \left( B^T (M ^T )^{-1}\Lambda \right. \\
\left. + \Lambda M^{-1} B \right )
\end{array} &
\begin{array}{c}  %23 array
\frac{\alpha}{2} \Lambda M^{-1} D \\
+ \frac{ \beta}{2} B^T (M ^T )^{-1} C^T L_{\mu}^{-1} A_{\mu} \\
+ \frac{\gamma}{2} \Lambda E L_{\mu}^{-1} A_{\mu}
\end{array} \\
\mbox{  }  \\
\begin{array}{c} %31 array
\left(\frac{\alpha}{2}(M^T )^{-1} (R-\tilde{ S})M^{-1}D\right.\\
 + \frac{\alpha}{2} M^{-1} D L_{\mu} ^{-1} A_{\mu} \\
 - \left.\frac{\gamma }{2} (M^T )^{-1} \Lambda E \right)^T  \\
\end{array} &
\begin{array}{c} %32 array
\left( \frac{\alpha}{2} \Lambda M^{-1} D \right. \\
+ \frac{ \beta}{2} B^T (M ^T )^{-1} C^T L_{\mu}^{-1} A_{\mu} \\
+ \left. \frac{\gamma}{2} \Lambda E L_{\mu}^{-1} A_{\mu} \right) ^T
\end{array} &
 \begin{array}{c} %33 array
\frac{\alpha}{2}  \left( D^TM^{-1} C^T L_{\mu }^{-1} A_{\mu}\right. \\
\left.- A_{\mu} L_{\mu}^{-1} C(M ^T )^{-1} D\right)
\end{array}
\end{array} \right]
\end{equation}

We now show that $ - \dot{W} $ is positive definite for suitable choices of
$ \alpha, \beta, \gamma,
B, D $, and $ E. $ To do this we block diagonalize $ -\dot{W} $ by repeated
applications
of lemma {\bf 2.11}.  Write $ - \dot{W} $ in its partitioned form as
\begin{equation}
- \dot{W} = \left[ \begin{array}{ccc} A_{11} & A_{12} & A_{13} \\
\\
A_{12}^T  & A_{22} & A_{23} \\
\\
 A_{13}^T & A_{23}^T & A_{33} \end{array}
\right] .
\end{equation}
Then
\[ \left[ \begin{array}{ccc} {\bf 1} & 0 & 0 \\
\\ -A_{12}^T A_{11}^{-1} & {\bf 1} &
0 \\
\\ 0 & 0 & {\bf 1} \end{array} \right]
\left[ \begin{array}{ccc} A_{11} & A_{12} & A_{13} \\
\\
A_{12}^T  & A_{22} & A_{23} \\
\\
 A_{13}^T & A_{23}^T & A_{33} \end{array}
\right] \left[ \begin{array}{ccc} {\bf 1}  & -A_{11}^{-1}A_{12} & 0 \\
\\
 0 & {\bf 1} & 0\\
\\
 0 & 0 & {\bf 1} \end{array} \right] \]
\begin{equation}  = \left[ \begin{array}{ccc} A_{11} & 0 & A_{13} \\
\\
 0 & A_{22}
- A_{12}^T A_{11}^{-1} A_{12}
 & -A_{12}^T A_{11}^{-1} A_{13} + A_{23} \\
\\
 A_{13}^T & -A_{13}^T A_{11}
^{-1} A_{12} + A_{23}^T & A_{33} \end{array} \right].
\end{equation}
Multiplying (3.11) by
\begin{equation} \left[ \begin{array}{ccc} {\bf 1} & 0 & -A_{11}^{-1} A_{13} \\
\\ 0
& {\bf 1} & 0 \\
\\ 0 & 0 & {\bf 1} \end{array} \right]
\end{equation}
on the right and by the transpose of (3.12) on the left yields
\begin{equation}
\left[ \begin{array}{ccc} A_{11} & 0 & 0 \\
\\ 0 & A_{22} -
A_{12}^T A_{11}^{-1} A_{12} & -A_{12}^T A_{11}^{-1} A_{13} + A_{23} \\
\\ 0 &
-A_{13}^T A_{11}^{-1} A_{12} + A_{23}^T & -A_{13}^T A_{11}^{-1} A_{13} + A_{33}
\end{array} \right].
\end{equation}
Then a final application of the lemma to (3.13) yields the block diagonal form
\begin{equation}
\left[ \begin{array}{ccc} A_{11} & 0 & 0 \\ 0 & A_{22} - A_{12}^T
A_{11}^{-1} A_{12} & 0 \\ 0 & 0 & \tilde{A}_{33} \end{array} \right]
\end{equation}
where
\begin{eqnarray}
\tilde{A}_{33} & = & A_{33} - A_{13}^T A _{11}^{-1} A_{13}
 - (A_{23}^T - A_{13} A _{11}^{-1} A_{12}) \times   \nonumber \\
&  & ( A_{22} - A_{12}^T A_{11}^{-1} A _{12} ) ^{-1} ( A_{23} - A_{12}^T A
_{11}^{-1} A _{13}) .
\end{eqnarray}
Now $ A_{11} $ in (3.10) is positive definite if $ \alpha$  and
$ \beta $ are small, since $ R $ is positive definite.  Choose, as in
Theorem {\bf 3.1},
$ B = M K ^{-1} \Lambda$, and assume that $\gamma$ is small, then
$A_{22}-A_{12}^TA_{11}^{-1}A_{12}=\beta\Lambda K^{-1}\Lambda-A_{12}^T
A_{11}^{-1}A_{12}.$
Since the second term is of higher order in $ \alpha, \beta, \gamma, $ this
is positive
definite. It remains to prove positive definiteness of $ \tilde{A}_{33}. $
Firstly, choose
\begin{equation}
D = M ^T K^{-1}C^TL_{\mu}^{-1}A_{\mu}:{\cal S}_{\rm RIG}\rightarrow{\cal W}
_{\rm INT}^{\ast}
\end{equation}
Then we find:
\begin{eqnarray*}
A_{11}& = & (M ^T) ^{-1}  R M^{-1}+ O(\alpha) + O(\beta) \\
A_{12} & = & \frac{ \beta }{ 2}(M ^T )^{-1} ( R - \tilde{S}) K ^{-1} \Lambda -
\frac{ \gamma }{2} (M ^T )^{-1} C^T L_{\mu} ^{-1} E^T \Lambda  \\
A_{13} &  = & \frac{ \alpha }{2} (M ^T )^{-1} (R - \tilde{S} )K ^{-1}  C^T
L_{\mu}
^{-1} A_{\mu} + \frac{ \alpha }{2} M ^{-1} M K ^{-1} C^T L_{\mu} ^{-1} A_{\mu}
L_{\mu} ^{-1} A_{\mu} \\
& & \mbox{} - \frac{ \gamma }{2} (M ^T )^{-1} \Lambda E \\
A_{23} & = & \left ( \frac{ \alpha}{2 } + \frac{ \beta }{2} \right) \Lambda
K ^{-1}  C^T L
_{\mu} ^{-1} A_{\mu} + \frac{ \gamma}{2} \Lambda E L_{\mu} ^{-1}A_{\mu} \\
A_{22} & = & \beta \Lambda K ^{-1} \Lambda  \\
A_{33} & = & \alpha (C ^T L _\mu ^{-1} A _\mu ) ^T K ^{-1} ( C ^T L _\mu
^{-1} A _\mu ).
\end{eqnarray*}
To show that $ \tilde{A}_{33} $ is positive definite, we set $ \alpha = \beta =
\gamma $ and write it as a term linear in $ \alpha $ plus higher order
terms in $ \alpha. $ Then we show the term linear in $
\alpha $ is indeed positive definite for $ C^T $ injective and a
suitable choice of $ E. $

We now isolate the terms in $ \tilde{A}_{33} $ that are linear in $\alpha.$
Since
\[ A_{11} = (M ^T )^{-1}R M^{-1} + O(\alpha ),\]
 we get $ A_{11}
^{-1} = M R^{-1} M ( 1 + O( \alpha)). $  Also $ A_{13}, A_{23} $ and $ A_{33} $
are all $ O(\alpha ). $  Hence
\[
A_{13}^T A_{11}^{-1} A_{13} = O( \alpha ^2)
\]
and so does
not affect definiteness, for small $ \alpha $.
Next,
\[
A_{22} - A_{12}^{T}A_{11}^{-1}A_{12} = \alpha \Lambda K ^{-1} \Lambda + O(
\alpha
^2) = \alpha \Lambda K ^{-1} \Lambda  ( 1 + O( \alpha) )
\]
and so
\[
(A_{22} - A_{12}^T A_{11}^{-1} A_{12} )^{-1} = \frac{ 1}{\alpha  }
\Lambda^{-1}K\Lambda^{-1}( 1 + O( \alpha )).
\]
Also,
\[
A_{23} - A _{ 12} ^T A _{ 11} ^{-1} A _{ 13} = A_{23} + O( \alpha ^2)
\]
and thus
\begin{eqnarray*}
\tilde{A}_{33} & = & A_{33} + O(\alpha ^2) - (A_{23}^T + O(\alpha ^2))
\frac{1}{\alpha } \Lambda^{-1}K\Lambda^{-1}( 1 + O( \alpha )) ( A_{23} + O(
\alpha ^2))
\nonumber  \\
& = & A_{33} + O(\alpha ^2 ) - \frac{ 1}{\alpha  } (A_{23}^T
\Lambda^{-1}K\Lambda^{-1}
+ O(\alpha ^2)  +  A_{23}^T \Lambda^{-1}K\Lambda^{-1}O( \alpha)\\
& & \mbox{} + O( \alpha ^3)) (A_{23} + O ( \alpha ^2 )) \nonumber  \\
& = &  A_{33} - \frac{ 1}{\alpha}A_{23} ^T\Lambda^{-1}K\Lambda^{-1}A_{23} + O(
\alpha ^2) .
\end{eqnarray*}
Hence the term in
$ \tilde{A}_{33} $
linear in $ \alpha $ is given by
\[
\alpha (C ^T L _\mu ^{-1} A _\mu ) ^T K ^{-1} (C ^T L _\mu ^{-1} A _\mu )
\hspace{3.2in}
\]
\[
+ \frac{1}{\alpha} (\alpha A_{\mu}L_{\mu}^{-1}C K ^{-1} \Lambda  +
\frac{1}{2} \alpha
A_{\mu}  L_{\mu}^{-1} E^T \Lambda  ) \Lambda^{-1}K\Lambda^{-1} ( \alpha
\Lambda K ^{-1}  C^T
L_{\mu}^{-1} A_{\mu} + \frac{1}{2} \alpha \Lambda E  L_{\mu}^{-1} A_{\mu} )
\]
\begin{eqnarray*}
& = & (C ^T L _\mu ^{-1} A _\mu ) ^T K ^{-1} (C ^T L _\mu ^{-1} A _\mu )  + \,
A
_\mu L _\mu ^{-1} C K ^{-1} C ^T L _\mu ^{-1} A _\mu \\
& & \quad + \, \frac{1}{2} \alpha
A_{\mu} L_{\mu}^{-1} E^T C^T  L_{\mu}^{-1} A_{\mu} + \frac{1}{2} \alpha
A_{\mu} L_{\mu}^{-1} C E L_{\mu}^{-1} A_{\mu} \\
& & \quad + \, \frac{1}{4} \alpha A_{\mu}
L_{\mu}^{-1} E^TKE  L_{\mu}^{-1} A_{\mu}.
\end{eqnarray*}
Since $ (A_{\mu} L_{\mu}^{-1}) = - (L_{\mu}^{-1} A_{\mu})^T, $ the first
two terms cancel and we obtain
\[
-\frac{\alpha}{2} ( L_{\mu}^{-1} A_{\mu})^T [ CE + (CE)^T + \frac{1}{2}
E^TK E ]( L_{\mu} ^{-1} A_{\mu}).
\]
Now let $ E = -K ^{-1} C^T.$  Then $ CE + (CE)^T + \frac{1}{2} E^TK E = -
\frac{3}{2}
CK ^{-1} C^T $ and hence
\[
\tilde{A}_{33} = \frac{ 3 \alpha }{4} (L_{\mu}^{-1} A_{\mu})^T
(CK ^{-1} C^T)(L_{\mu}^{-1} A_{\mu}) + O(\alpha ^2) \]
which is positive definite since $ C ^T $ is injective and $ \alpha > 0. $

Since $W$ is clearly indefinite and $ \dot{W} $ is negative
definite, we have Lyapunov instability by Lyapunov's instability theorem.

To prove the theorem in the case that $\Lambda$ is degenerate, split the
variable
$q$ into $( q_1  , q _2 )$ as in the proof of the abelian case and note that
the
equations (3.2) decouple into equations for $ q _1 $ and $ (r, p, q _2)$.
Now repeat
the argument using the same modifications as in the abelian case.

For completeness, we give the details in the extreme case $ \Lambda = 0 $.
In this case, the linearized dynamics with added dissipation in the
block-diagonal normal form takes the following ``triangular'' form:
\begin{equation}
\left. \begin{array}{lcl}
M\dot{q}& = & p \\
\\
\dot{p}
&=& -(\tilde{S}+R) M^{-1}p-C^T L_\mu^{-1} A _\mu r\\
\\
\dot{r}
&=& -L_\mu ^{-1} A _\mu ^{-1}r-L_\mu ^{-1}C M ^{-1}  p, \end{array}
\right\} \end{equation}
where $\tilde{S} = S + C^T L_\mu ^{-1} C, $ and $  R =
R^T \geq 0 $ is a matrix of damping coefficients.  Note that projecting out the
shape variable $q$ leaves the reduced system from (3.18) involving $
\dot{p} $ and $ \dot{r} $
only, which can be handled separately. Let
\[ W(p, r) = \frac{1}{2} p \cdot M^{-1} p
+ \frac{1}{2} r \cdot A_\mu r + \alpha D r \cdot M^{-1} p \]
where $ \alpha $ is a scalar and $ D $ is to
be chosen.  We will show that $ \alpha $ and $ D $  can be so chosen that $
W(p, r) $ is a Chetaev function for the {\it reduced\/} system---the second and
third equations of (3.18) {\it i.e.,\/} $W(p, r)$ is indefinite and its
total derivative $
\dot{W} $ along trajectories of (3.18) is negative definite.  This would
then establish
the Lyapunov instability of the reduced system and consequently of
the full system (3.18). As above, choose
\[ D = MK ^{-1}  C^T L^{-1} _\mu A_\mu.\]
It is then easy to verify that
\[
\dot{W} = - (p^T \,r^T)
 \left( \begin{array}{cc} Q_{11}
& Q _{12} \\
\\
 Q^T_{12} & Q_{22}\end{array} \right) \left( \begin{array}{c}  p \\ r
\end{array}
\right)
\]
where,
\begin{eqnarray*}
Q_{11} & = & (M ^T )^{-1} R M^{-1} + O(\alpha ) \\
\\
Q_{12} & = & \frac{ \alpha }{ 2}\left( (M^T )^{-1}
(R-\tilde{S})K ^{-1} C^T L^{-1}_\mu
A_\mu + K ^{-1} C^T  L^{-1}_\mu A_\mu L^{-1}_\mu A_\mu \right) \\
\\
Q_{22} & = &\alpha (C ^T L _\mu ^{-1} A _\mu ) ^T K ^{-1} ( C ^T L _\mu
^{-1} A _\mu ).
\end{eqnarray*}
By hypothesis $Q_{22} > 0.$ As above, there
is a range of $ \alpha $ for which the matrix
\[
Q _{\alpha} =  \left( \begin{array}{cc} Q_{11}
& Q _{12} \\
\\
Q^T_{12} &
Q_{22}\end{array} \right)
\]
is positive definite.  Further, since the signature of a hyperbolic matrix is
invariant under small perturbations, one can further choose $ \alpha $ in
the range $
(0, c) $ such that,
\[
\mbox{signature} \left( \begin{array}{cc} M^{-1}
& \displaystyle{\frac{\alpha }{2}} M^{-1} D  \\
\\
 \displaystyle{\frac{\alpha }{2}} D ^T M^{-1} &
A_\mu \end{array} \right) = \mbox{signature} \left( \begin{array}{cc} M^{-1}
& 0 \\ 0 &
A_\mu \end{array} \right).
\]
The matrix $ \left( \begin{array}{cc} M^{-1}
& 0 \\ 0 &
A_\mu \end{array} \right) $ is indefinite by hypothesis.  Thus we have a range
of $ \alpha $ for which $ W $ is a Chetaev function and we have proved
Lyapunov instability. \quad $\blacksquare$
\vspace{0.2in}

\noindent{\large \bf Remark\,} We leave it to the reader to verify that
standard
eigenvalue inequalities lead to the condition,
\[0 < \alpha < c, \]
where
\begin{eqnarray*}
c & = &  \min\{ c_1, c_2^{-1}\}, \\
c_1 & = &  \left\{ \begin{array}{lcc} \infty
& \mbox{if} & \lambda_{{\rm min}} ( C^T L_\mu ^{-1} A_\mu L_\mu ^{-1} C)
\geq 0 \\  \displaystyle{\frac{\lambda_{{\rm min}} ( M^{-1} R
M^{-1})}{|\lambda_{{\rm min}}
 (C^T L_\mu ^{-1} A_\mu L_\mu ^{-1} C)| }} & \mbox{if} & \lambda_{{\rm min}}
 ( C^T L_\mu ^{-1} A_\mu L_\mu ^{-1} C)
< 0,
\end{array} \right.\\
c_2 & = & ||(M ^{-1} R M ^{-1} )^{-1/2} C^T L^{-1}_\mu A_\mu L_\mu ^{-1} C ( M
^{-1} R M ^{-1} )^{-1/2} || \\
& & + \lambda _{{\rm max}} (Q^T_{12} MR ^{-1}
 M Q_{12} ) / \lambda _{{\rm min}}(Q _{22}) \end{eqnarray*}
and $ ||\cdot || $ denotes the Euclidean norm. $\quad \blacklozenge $
\vspace{0.2in}

An illustration of the instability result of Theorem
{\bf 3.3} in the case of $\Lambda = 0 $, and of the effective use of the
block diagonal
normal form will be given in the examples in \S {\bf 6}.

\section{Instability of Relative Equilibria}

Our main result shows that if $ \delta^2 H_{\xi} $ is indefinite at a
given relative equilibrium, the system is {\tenbi dissipation unstable\/} about
that equilibrium.  To do this, it is sufficient to prove {\tenbi spectral\/}
instability of the linear system (3.2). In \S {\bf 3} we proved Lyapunov
instability of
this system. As discussed in the introduction, this is not sufficient to prove
instability of the nonlinear system.  Hence we need to show that we do in fact
have spectral instability.  This will follow from the following proposition
which utilizes the eigenstructure of the linearized Hamiltonian  system ({\it
i.e.,\/} with $ R = 0 $) and a key observation of Hahn [1967].

\begin{prop} Let $ \dot{x} = X_H(x) $ be a linear Hamiltonian system.
Suppose that one adds a small linear perturbation to $ X_H $ (in
particular, a damping term) and that for the augmented system there exists
a quadratic  form $W$ which has at least one negative eigendirection and
which satisfies $ \dot{W} < 0 $ along the flow.  Then the augmented system
is spectrally unstable.\end{prop}

\noindent{\bf Proof} \quad The properties of $W$ imply that the augmented
system is Lyapunov unstable, as we have seen.  We now show that it is
spectrally unstable.  Henceforth, we shall refer to the augmented system
as the damped system and the perturbation as damping.

We consider firstly the eigenvalue configurations of the undamped linear
Hamiltonian system.  From the general properties of Hamiltonian matrices
(see e.g. Abraham and Marsden [1978]) the possible configurations can be
grouped
into the following four categories:
\begin{enumerate}
\item  There is at least one quadruplet, {\it i.e.,\/} an eigenvalue
configuration shown in Figure 4.1:

\begin{figure}%notes
\vspace{2.1in}
\hspace*{.8in} \special{illustration fig4.1.eps}
\caption{The case of an eigenvalue quadruplet.}
\end{figure}

\item There is at least one pair of real eigenvalues, as in Figure 4.2.

\begin{figure}%notes
\vspace{2.1in}
\hspace*{.8in} \special{illustration fig4.2.eps}
\caption{The case of eigenvalues on the real axis.}
\end{figure}

\item Neither 1. nor 2. holds but all the eigenvalues are on the imaginary
axis and are simple.

\item  All the eigenvalues are on the imaginary axis and there is at
least one multiple eigenvalue. \end{enumerate}

Now add the damping terms.  In cases 1. and 2., small damping leaves
eigenvalues in the right half plane.  Hence we have spectral
instability.  Now consider case 3.  All eigenvalues cannot move to the
left half plane since this implies Lyapunov stability and we have
instability.  They cannot all remain on the imaginary axis since (for
small damping) they remain distinct and hence all solutions would be
periodic and hence stable.  Similarly if some move into the left half
plane and some remain (distinct) on the imaginary axis, the system is still
stable.  The only remaining possibility is at least one moves to the
right half plane and we thus have spectral instability.

Finally consider case 4.  If any eigenvalues move into the right half
plane we have spectral instability and are done.  Now if all eigenvalues
moved to the left half plane the system would be stable and we know it is
unstable.  Similarly it is impossible for some to move to the left and
for those that remain on the imaginary axis to be simple, for this again
implies stability.

The only remaining possibilities are a multiple zero eigenvalue or a
multiple pair of conjugate purely imaginary eigenvalues remaining on the
imaginary axis after the addition of damping.  We can show that both
situations are impossible for they contradict $ \dot{W} < 0: $

Suppose firstly that there is a zero eigenvalue.  Let $ W = z^T Q z$  and
$  X_{H}(z) = Az. $ Then $ \dot{W} = z^T (A^T Q + QA)z.$   But there exists an
$ \tilde{z} \neq 0 $ and that $ A \tilde{z} = 0 $ and here $ \dot{W}
(\tilde{z}) = 0, $ contradicting $ \dot{W} < 0. $

Now suppose there is a pair of conjugate purely imaginary multiple
eigenvalues.  Then there exists an invariant subspace for the flow, which
is a subspace of the generalized eigenspace corresponding to the multiple
eigenvalues, which is invariant for the matrix
\begin{equation}
\left[ \begin{array}{cccc} 0 & -b & 0 & 0 \\ b & 0 & 0 & 0 \\
1 & 0 & 0 & - b \\ 0 & 1 & b & 0 \end{array} \right] .
\end{equation}

Now we can use the following argument of Hahn [1967].  There exists a
solution of
the system corresponding to (4.1) of the form
\[z_1 = z_2 = 0 \quad \quad z_3 =
\cos b t \quad \quad z_4 = {\rm sin}\, bt. \]
However, $W$  is a periodic function of $t$ when evaluated on the above
solution.  On the other hand,
\begin{equation}
W (t) = W_0  + \int_{t_0}^t \dot{W}(s) ds.
\end{equation}
Since $ \dot{W} < 0 $, $ |\dot{W}| $ is bounded away from
zero on this curve.  Hence the integral would not be bounded as $ t
\rightarrow \infty $ and $W$ cannot be periodic on this trajectory.
Hence we cannot have a pair of conjugate purely imaginary multiple
eigenvalues since this contradicts  $ \dot{W} < 0. $

Thus we see that at least one eigenvalue must be in the right half plane
and the system is spectrally unstable. \quad $\blacksquare$

 \vspace{0.2in}

Combining our results and using the notation of \S{\bf 2}, we get

\begin{thm} Assume (for non-abelian groups) $ C^T $ is injective, and the
second variation of the
Energy-Momentum function $ H_{\xi} $ of the Hamiltonian system is
indefinite at a given relative
equilibrium.  Then the addition of strong (internal) Rayleigh dissipation
gives spectral instability
of the system about that relative equilibrium.  \end{thm}

The arguments we have given are designed especially to be applicable to
infinite dimensional systems, even though we have so far confined our
attention to finite dimensional ones.  There need to be appropriate
assumptions on the semigroups involved, and assumptions on the spectra,
but it seems that the main assumption needed for the above analysis to be
valid is that the spectrum of the unperturbed problem be discrete, with
eigenvalues having at most finite multiplicity.

Two interesting problems are the whirling string and the rotating
circular liquid drop.  We hope to pursue the analysis of these problems
using the present techniques in another publication. We analyze a simple
rotating beam, where the
infinite-dimensional calculation reduces to a finite-dimensional one, in \S
{\bf 6}.

\vspace{0.2in}

We now make some remarks on the condition requiring $ C^T $ to be injective.
\vspace{0.2in}

\noindent{\large \bf Remarks\,}

{\bf 1} Since $ C ^T : {\cal V} _{\rm RIG} \rightarrow {\cal W} _{\rm INT}
^{\ast} $, note
it can be injective only if $ {\rm dim}\,{\cal V} _{\rm RIG} = {\rm
dim}\,\frak g _{ \mu
_e} ^{\perp} \leq {\rm dim}\,Q - {\rm dim} \, G$, {\it i.e.,\/} $ {\rm
dim}\,{\cal O} _{
\mu _e} \leq {\rm dim}(Q/G).$ For example, for $SO(3)$, this says that $ 5
\leq {\rm dim}\,
Q $. For a rigid body with rotors and $ G = SO(3)$, this says that there
must be at least
two rotors.

{\bf 2} We claim that $ C ^T $ is injective, {\it i.e.,\/} $C$ is surjective,
if
\[
{\cal V} _{\rm INT} ^{\perp} \cap \frak g \cdot q _e = \{0\}.
\]
\noindent {\bf Proof\,} From (2.34),
\[
\left\langle C ^T(\Delta q), \delta q \right\rangle = - \langle \! \langle
\zeta _Q(q _e ), \delta q \rangle \! \rangle .
\]
Suppose this is zero for all $ \delta q \in {\cal V} _{\rm INT} $. Then $
\zeta _Q(q _e) \in
{\cal V} _{\rm INT} ^{\perp} $ and so by hypothesis, $ \zeta _Q(q _e) = 0
$. By freeness, $ \zeta =
0$, and so by Lemma {\bf 2.7}, $ \zeta = {\Bbb I} (q _e ) ^{-1} {\rm
ad}_\eta ^{\ast} \mu$, where $
\eta \in \frak g _{ \mu _e} $, and so as $ \eta \in \frak g _{ \mu _e}
^{\perp} $, by (2.22), $
\eta = 0 $, and so by (2.32), $ \Delta q = 0. $ \quad $\blacksquare$

\vspace{0.2in}

Notice that the above condition is a hypothesis on $ {\cal V} _{\rm INT} $
being
``genuinely different'' from the ``naive'' choice of internal space, namely
$ [\frak g \cdot q _e ] ^{\perp} $.
\vspace{0.2in}

\noindent{\bf Remark on Internal Symmetries}:  In some situations, we will
have internal symmetries in the system.  In the
Hamiltonian case each such symmetry would enable one to reduce the system
by one degree of freedom.  In the presence of damping (dissipation) we
cannot of course do this, but one can nonetheless eliminate the
corresponding configuration variables.  This ensures that the matrix
representation of $ \dot{W} $ will be negative semi-definite rather than
definite (with zero eigenvalues due to the symmetry).  The same
analysis as before then applies.  This situation will be illustrated in \S
{\bf 6}.

\section{Dissipation-Induced Movement of Eigenvalues}
In contrast with the method of Routh-Hurwitz that requires explicit
calculations with characteristic polynomials, the methods of the present paper
allow one to predict dissipation-induced instability of relative equilibria
solely on the basis of signature computations---indefiniteness of $ \delta
^2H_{\xi} $ and injectivity of $C^T. $ In this sense, the present
paper is closer in spirit to the work of Hermite on Hankel quadratic forms,
{\it cf.\/} the last chapter of vol 2. of Gantmacher [1959].  However, the
classical work of Routh,
Hermite and Hurwitz was aimed at getting more refined information---such as
the {\it number\/} of
right half plane eigenvalues---than just predicting instability.  In the
present context, a
closely related question is that of determining speeds of crossing (into the
right half plane) of pure imaginary eigenvalues due to dissipative
perturbations of an underlying Hamiltonian system.  In this section we discuss
some formulae to compute such speeds and thereby track in detail the mechanism
of instability.  Our formulae generalize the previous work of Krein [1950]
and McKay
[1991], and when specialized to the block-diagonal normal form (abelian as
well as non-abelian cases) yield new and explicit formulae for crossing
speeds.

Keeping in mind the well-known connections between the asymptotic stability
of a linear system and solutions to the matrix Lyapunov equation {\it cf.\/}
Bellman [1963], Brockett [1970], Taussky [1961], our proofs will have  a
definite
Lyapunov theory flavor.  In particular, we will not need the Kato perturbation
lemma {\it cf.\/}  McKay [1991].  We first prove a basic result.

\begin{lem} Consider a linear system $ \dot{x} = A x $ and a quadratic form $
V(x) = \frac{1}{2} x^T Qx. $ Let $ \dot{V} (x) $ denote the total derivative of
$ V $ along trajectories of the linear system, evaluated at $ x. $ Let $
\lambda
= \lambda_r + i \lambda_i \in {\rm spectrum}(A).$ Let $ \xi = x_r + i x_i $
denote an eigenvector of $A$  corresponding to $\lambda$. Then, \end{lem}
\begin{equation} \lambda _r = \frac{\dot{V} (x_r) + \dot{V} (x_i)}{2 ( V ( x_r)
+ V ( x_i))} .
\end{equation}
\noindent {\bf Proof} \quad
\begin{eqnarray*}
\dot{V} (x) & = & \frac{1}{2} (\dot{x} ^T Q x + x^T
Q\dot{x}) = \frac{1}{2} x^T (A^T Q + QA) x, \quad  \mbox{and thus} \\
\dot{V} (\xi) & = & \frac{1}{2} \xi ^T (A^TQ + QA) \xi =
\frac{1}{2} 2 \lambda \xi ^T Q \xi = 2 \lambda V (\xi).
\end{eqnarray*}
Thus
\[ \lambda = \displaystyle{\frac{\dot{V} (\xi)}{V(\xi)}}. \]
Now
\begin{eqnarray}
 \frac{\dot{V}(x_r)}{V (x_r)} & = & \frac{x^T_r ( A^TQ + Q A) x_r}{x_r^T Q x_r}
 \nonumber\\ & = & \frac{(\lambda _r x_r^T - \lambda_i x_i^T) Q x_r + x_r ^T Q
(
\lambda_r x_r - \lambda_i x_i)}{x_r^T Q x_r}  \nonumber\\ & = & 2 \lambda_r - 2
\lambda_i \frac{x_i^T Q x_r}{x_r ^T Q x_r}. \end{eqnarray}
Similarly,
\begin{equation}
\frac{\dot{V}(x_i)}{V(x_i)} = 2 \lambda_r + 2 \lambda _i \frac{x_i^T Q
x_r}{x_i^T Q x_i} . \end{equation}
Adding suitable multiples of (5.2) and (5.3) we get,
\begin{eqnarray*}
2 \lambda  _r ( x_r^T Qx_r + x_i^T Qx_i)
& = & \frac{\dot{V}(x_r)}{V(x_r)} x_r^T Qx _r + \frac{\dot{V}(x_i)}{V(x_i)}
x_i^T Qx_i \\ & = & 2(\dot{V} (x_r) + \dot{V} (x_i)).
\end{eqnarray*}
Therefore,
\[ \lambda_r = \frac{\dot{V} (x_r) + \dot{V}(x_i)}{x_r^T Qx_r + x_i^T Qx_i} =
\frac{\dot{V} (x_r) + \dot{V} (x_i)}{2(V(x_r) + V(x_i))}. \quad \blacksquare
\]

\vspace{0.2in}

\begin{cor}
Consider the matrix Lyapunov equation
\begin{equation}
A^T X +  XA = -P
\end{equation}
associated to the linear system $ \dot{x} = Ax, $ where $ P = P^T > 0 $ is
given.  Suppose $ Q = Q^T $ is a solution to {\rm (5.4)}.  Then,
\begin{equation}
 \mbox{{\rm card}}\{ \lambda \mid  \lambda \in \,\mbox{{\rm spectrum}}(A),
{\rm Re}
(\lambda) = \lambda _r > 0 \}
 \leq \,\mbox{{\rm index}}(Q),
\end{equation}
where $ {\rm index}(Q) $ means the number of negative eigenvalues of $Q$.
\end{cor}
\noindent {\bf Proof} \quad In Lemma {\bf 5.1}, choose $ Q $ to be a solution
to the Lyapunov equation (5.4).  Then,
\begin{eqnarray*}
 \dot{V} (x) & = & \frac{1}{2} (\dot{x}^TQx + x^TQ\dot{x}) \\
& = & \frac{1}{2} x^T (A^TQ + QA)x \\
& = & - \frac{1}{2} x^T Px.
\end{eqnarray*}
 From Lemma {\bf 5.1}, for any eigenvalue $ \lambda$  of $A$,
\begin{eqnarray*}
\lambda_r & = & \frac{\dot{V}(x_r) + \dot{V} (x_i)}{2(V(x_r) + V (x_i)} \\
& = & \frac{ - \frac{1}{2}(x_r^T Px_r + x_i^T Px_i)}{(x_r^T Q x_r + x_i^T Q
x_i)} \\ & = & -\frac{1}{2} \frac{\bar{\xi}{} ^T P \xi}{\bar{\xi}{} ^TQ \xi} .
\end{eqnarray*}

\noindent Since $ P = P^T > 0,\, \lambda_r > 0,$  this implies $\bar{\xi}{}{}^T
Q\xi < 0. $  \quad $\blacksquare$
\vspace{0.2in}

\noindent{\bf Remark} \quad It is well-known that when spectrum($A$) lies in
the
strict left half plane, (5.4) has the unique positive definite solution

\[ \int_0^{\infty} e^{A^{T }\sigma} Q e ^{A \sigma} d \sigma. \]
 If in Corollary
{\bf 5.2}, we impose the  additional condition that, for any $ \lambda, \mu
\in$  spectrum$(A), \lambda + \mu  \neq 0, $ then, the inequality (5.5)
becomes an equality.  This is a theorem of  Taussky [1961]. \quad
$\blacklozenge$
\vspace{0.2in}

Suppose the linear system of interest is

\begin{equation} \dot{x} = [\Omega ^{-1} ]^T Qx + \epsilon Bx
\end{equation}
where $ \Omega = - \Omega ^T $ is a nonsingular matrix ({\it e.g.\/} the
symplectic structure) of size $ 2n \times 2n, B $ determines a, possibly
dissipative, perturbation, $ \epsilon \geq 0 $ is a small parameter, and $ Q =
Q^T $ determines the energy quadratic form
\begin{equation}
E(x) = \frac{1}{2} x^T Qx
\end{equation}
for the underlying unperturbed system.  Along trajectories of (5.6)
\begin{equation}
\dot{E} (x) = \,\epsilon  x^T QBx.
\end{equation}

\begin{cor}
Suppose $ \lambda $ is a simple eigenvalue of $ A = \Omega ^{-T} Q $ with
eigenvector $ \xi = x_r + ix_i. $ Let $ \lambda _r^\epsilon $ denote the real
part of the eigenvalue branch $ \lambda ^\epsilon $ of $ A_\epsilon = \Omega
^{-T} Q + \epsilon B $ emanating from $ \lambda. $ Then
\begin{eqnarray}
\lambda _r^\prime :&= & \left.\frac{d}{d \epsilon } \lambda _r^\epsilon
\right|_{\epsilon = 0}  \nonumber\\
& = & \frac{1}{2} \frac{\dot{E} ^\prime (x_r) + \dot{E}^\prime (x_i)}{E(x_r) +
E(x_i)}
\end{eqnarray}
 where
\begin{equation}
\dot{E} ^\prime (x) = x^T Q Bx.
\end{equation}
\end{cor}
\noindent {\bf Proof} \quad Substitute $ E(x) $ for $ V(x) $ in Lemma {\bf 5.1}
and observe that simplicity of $\lambda$ ensures smoothness of $ \dot{E}
(x_r^\epsilon ), \dot{E} (x_i^\epsilon ) $ etc. with respect to $\epsilon$ at $
\epsilon = 0. $ \quad $\blacksquare$

\vspace{0.2in}
If in Corollary {\bf 5.3}, the eigenvalue branch $ \lambda ^\epsilon $ is
emanating from a pure imaginary eigenvalue $ \lambda = i \omega, $ then the
formula (5.9) becomes a formula for the {\it crossing speed}. It is our aim to
make this formula explicit for systems in block-diagonal normal form.  As a
first
step we note
\begin{lem} Under the hypotheses of Corollary {\bf 5.3}, and if $ \lambda =
i \omega $
where $ \omega \in {\Bbb R} $,
\begin{equation}
\lambda ^\prime _r = \frac{\bar{\xi}{} ^T (\Omega B) _{{\rm anti}}\, \xi
}{\bar{\xi}{}^T \Omega \xi },
\end{equation}
where $ (\, \cdot\, )_{{\rm anti}} $ denotes the anti-symmetric part. \end{lem}

\noindent {\bf Proof} \quad Since $ A(x_r + ix_i) = i \omega (x_r + ix_i),$
we have
\begin{eqnarray*}
Ax_r  =  - \omega x_i  =   \frac{i\omega}{2} (\xi - \bar{\xi }),
\end{eqnarray*}
and
\[ Ax_i  =  \omega x_r = \frac{\omega }{2} (\xi + \bar{\xi}).\]
Then,
\begin{eqnarray}
E(x_r) & = & \frac{1}{2} x_r^T Q x_r  =  \frac{1}{2} x^T_r \Omega ^T A x_r
\nonumber\\
& = & \frac{1}{2} \left( \frac{\xi + \bar{\xi}}{2} \right)^T \Omega ^T \frac{i
\omega }{2} ( \xi - \bar{\xi} ) \nonumber\\
& = & - \frac{1}{4}  i \omega  \bar{\xi}\, ^T \Omega \xi.
\end{eqnarray}
Similiarly,
\begin{eqnarray}
E(x_i) & = & \frac{1}{2} x_i^T Q x_i  =  \frac{1}{2} x_i^T \Omega ^T A x _i
\nonumber\\
& = & \frac{1}{2} \left( \frac{\xi - \bar{\xi}}{2i}\right) ^T \Omega ^T
\frac{\omega }{2}  ( \xi + \bar{\xi})\nonumber\\
 & = & - \frac{1}{4}  i \omega \bar {\xi}\, ^T \Omega \xi.
\end{eqnarray}
Further,
\begin{eqnarray}
\dot{E}^\prime (x_r) + \dot{E}^\prime (x_i) & = & x_r^T Q Bx_r + x_i^T Q
Bx_i \nonumber\\
& = & x_r^T A^T \Omega B x_r + x_i ^T A^T \Omega B x_i \nonumber\\
& = & \frac{i \omega }{2} ( \xi - \bar{\xi} )^T \Omega B \left( \frac{\xi +
\bar{\xi}}{2}\right) \nonumber\\
& & + \frac{\omega }{2} ( \xi + \bar{\xi} )^T \Omega B \left( \frac{\xi -
\bar{\xi}}{2}\right) \nonumber \\
& = & - i \omega \bar{\xi}\,^T \left( \frac{\Omega B - ( \Omega B )^T}{2}
\right) \xi \nonumber\\ & = & - i \omega \bar{\xi}\,^T ( \Omega B) _{{\rm
anti}}
\xi.  \end{eqnarray}
 From (5.12), (5.13) and (5.14) we get,
\[
\lambda ^\prime_r  =  \frac{\dot{E} ^\prime (x_r ) + \dot{E}^\prime ( x_i)}{2
(E ( x_r) + E ( x_i))}   =  \frac{\bar{\xi}\,^T ( \Omega B ) _{{\rm
anti}} \xi}{\bar{\xi} \,^T \Omega \xi }.  \quad \blacksquare \]

\noindent{\bf Remark}\quad  The special case $ \Omega = J = \left(
\begin{array}{
cc} 0 & I \\ -I & 0 \end{array} \right) $ of formula (5.11) appears in R. McKay
[1991] who also gives it an averaging interpretation.  We note that our result
is a corollary of the more general formula (5.9) which applies to eigenvalues
that are not necessarily on the imaginary axis.  The proof presented here does
not use the Kato perturbation lemma involving both right and left
eigenvectors---the key tool in McKay's argument. \quad $\blacklozenge$
\vspace{0.2in}

Next we compute the average $ \langle \dot{E}^\prime (x_r) \rangle $ over a
cycle of period $ 2 \pi /\omega $ of the periodic solution $ \xi e^{i
\omega t} $ for the unperturbed system.
Recall that at $ t = 0, $ the formula
\[ \dot{E}^\prime (x_r) = \frac{i \omega }{4} (\xi - \bar{\xi} )^T \Omega B (
\xi + \bar{\xi }) \]
holds.  For any other $ t, $
\begin{eqnarray}
\dot{E}^\prime (x_r(t))& = & \frac{i \omega }{4} ( \xi e^{i \omega t} -
\bar{\xi} e
^{-i \omega t} ) ^T \Omega B ( \xi e ^{i \omega t} + \bar{\xi} e ^{- i \omega
t}
)  \nonumber\\ & = &\frac{i \omega }{4} \{ \xi ^T \Omega B e ^{i 2 \omega t} -
\bar{\xi}\, ^T \Omega B \xi + \xi{} ^T \Omega B \bar{\xi} - \bar{\xi} ^T
\Omega B
\xi e ^{-2 i \omega t} \}
\end{eqnarray}
Substituting from (5.15) into the average defined by
\begin{equation}
\langle \dot{E}^\prime (x_r) \rangle := \displaystyle{\frac{1}{ 2 \pi/\omega }
}
\int _0 ^{2 \pi / \omega} \dot{E} ^\prime (x_r (t)) dt  \end{equation}
we get,
\begin{equation}
\langle \dot{E}^\prime (x_r) \rangle  =  \frac{i \omega }{4} ( \xi{} ^T \Omega
B \bar{\xi} - \bar{\xi}{}^T \Omega B \xi)
 = - \frac{i \omega }{4} ( \bar{\xi}{}^T (\Omega B) _{{\rm anti}} \xi ) .
\end{equation}
In evaluating (5.16) we used the fact that $ \int _0 ^{2 \pi /\omega } e ^{i k
\omega t} dt = 0 $ for any nonzero integer $ k. $  From (5.14) and (5.11),
and (5.12),
(5.13) we get,
\begin{equation}
\lambda ^\prime _r = \frac{\langle \dot{E} ^\prime (x_r) \rangle}{2 E (x_r)}.
\end{equation}
This is the averaging interpretation of the crossing speed given by McKay in
the case  $ \Omega = J. $
\vspace{0.2in}

In the remainder of this section we show how to adapt the crossing-speed result
(5.11) to the block-diagonal normal form.  Recall that the symplectic
structure of the block-diagonal normal form is not canonical.  It is of the
form
``coadjoint orbit,  internal symplectic, magnetic and coupling terms'';
\begin{equation}
 \Omega = \left[ \begin{array}{ccc} L_\mu  & C & 0 \\ -C^T & S & {\bf 1} \\
0 & -
{\bf 1} & 0 \end{array} \right].
\end{equation}
The second variation $ \delta ^2 H_{\xi} $ takes the form,
\begin{equation} Q = \left[ \begin{array}{ccc} A_{\mu} & 0 & 0 \\ 0 & \Lambda
&
0\\ 0 & 0 & M^{-1} \end{array} \right],
\end{equation}
and the dissipatively perturbed linear system of interest is ({\it cf.\/}
equation (3.2)) \begin{equation}
\dot{x}  =  (A + \epsilon B) x \nonumber  =  ( \Omega ^{-T} Q + \epsilon B) x
\end{equation}
where
\begin{equation}
B = \left[ \begin{array}{cccc} 0 &0 & 0 \\ 0 & 0 & 0 \\ 0 & 0 & - RM ^{-1}
\end{array} \right].
\end{equation}
Here $ x = (r, q, p), $ as in \S {\bf 3}. A key stumbling block in using the
crossing speed formula (5.11) is the need to calculate the eigenvector $ \xi $
corresponding to the eigenvalue $ i \omega. $ The following result eases the
way a little.

\begin{lem}
Consider the quadratic pencil
\begin{equation} G (\lambda) =
\left[ \begin{array}{cc} \lambda ^2 M + \lambda S + \Lambda  & - \lambda C^T \\
\\
- \lambda L^{-T}_\mu C & \lambda {\bf 1}  - L_\mu ^{-T} C \end{array} \right].
\end{equation}
Then $ \lambda_0 $ is a singular point of the pencil, {\it i.e.,\/}  det $
[G(\lambda_0) ] = 0, $ with corresponding {\it null-vector\/} $ y_0 = (q^T_0,
r^T_0) ^T $ iff $ \lambda_0 $ is an eigenvalue of $ A $ with eigenvector $
\xi = ( r_0^T, q_0^T, \lambda_0 (Mq_0)^T)^T. $ \end{lem}

\noindent {\bf Proof}\quad  Note the equivalence between the unperturbed
system $
\dot{x} = Ax $ and the coupled system consisting of the second-order
internal dynamics
together with the first order coadjoint orbit dynamics, in normal form:
\begin{eqnarray}
M\ddot{q} + S \dot{q} + \Lambda q & = & C^T \dot{r}  \nonumber\\
\dot{r} & = & L_\mu ^{-T} A_\mu r + L_\mu ^{-T} C \dot{q}.
\end{eqnarray}
Interpret $ G(\lambda) $ as the Laplace transform representation of (5.24).
This immediately identifies singular points of the pencil $ G(\lambda)$ with
the spectrum of $A$.  The eigenvector-null vector result is a direct
calculation. \quad $\blacksquare$

\vspace{0.2in}
Now, suppose $ \lambda = i \omega _0 $ is a pure imaginary eigenvalue of $A$
(singular point of $ G(\lambda)).$  Let

\[G(i \omega _0) \left( \begin{array}{c}  q_0\\ r_0\end{array}
\right) = 0.\]
By Lemma {\bf 5.5}, and verifying that
\[ (\Omega B) _{{\rm anti}} = \frac{1}{2} \left[ \begin{array}{ccc } 0 & 0 & 0
\\ 0 & 0 & -RM^{-1} \\ 0 & RM^{-1} & 0 \end{array} \right], \]
we get
\begin{equation}
\bar{\xi}\,^T (\Omega B)_{{\rm anti}} \xi = - i\omega  _0 \,\bar{q}^T_0 Rq_0.
\end{equation}
Again by Lemma {\bf 5.5} and (5.19),
\begin{equation}
\bar{\xi}\,^T \Omega \xi  = \bar{r}^T_0 L_\mu r_0 + 2 i \omega _0 \,\bar{q}
^T_0
M \bar{q}_0
 +  \bar{q} ^T_0 S q_0
 +  \bar{r} ^T_0C q_0 -  \bar{q} ^T_0 C r_0.
\end{equation}
Setting,
\[ q_0 = \eta + i \beta, \]
\[ r_0 = u + iv \]
and substituting in (5.25), (5.26) we get the following ``block-diagonal''
version of the crossing speed formula,
\begin{equation}
\lambda ^\prime _r = \frac{- \omega_0 ( \eta ^T R \eta + \beta ^T R \beta)}
{2 \{ u^T  L_\mu v + u^T C \beta - v^T C \eta + \eta ^T S \beta + \omega_0 (
\eta ^T M \eta + \beta ^T M \beta ) \}} . \end{equation}

\noindent{\bf Remark}\quad  The crossing speed formula (5.27) can lead to
effective computation provided one has some insight into det$(G(i \omega _0)) $
and can compute a null vector of $ G(i \omega_0). $ This is still more
manageable
than directly computing eigenvectors of $ A $ due to  the smaller matrices
involved.

\vspace{0.2in}

\noindent{\bf Remark} \quad In the {\it abelian case\/}, $ L_\mu = 0 = C $ and
the crossing speed formula (5.27) specializes to
\begin{equation}
\lambda ^\prime _r = \frac{- \omega_0 ( \eta^T R \eta + \beta ^T R \beta) }{2
\{ \omega_0 ( \eta ^T M \eta + \beta ^T M \beta ) + \eta ^T S \beta \}}.
\end{equation}
For a similar formula for two degree of freedom systems, see Haller [1992].
Further, if there is no gyroscopic/magnetic term, {\it i.e.,\/} $
S = 0 $ then (5.28) predicts that every pure imaginary eigenvalue of the
unperturbed system is pushed into the {\it left half plane\/} under a strong
dissipation $ R >0. $ Of course, this says nothing about any eigenvalues of the
unperturbed system that may be in the right half plane---such eigenvalues are
bound to be present if $ S = 0 $ and $ \Lambda $ is indefinite.
\vspace{0.2in}

\noindent{\bf Example} As we already saw in the introductory section, for
the two
degrees of freedom Chetaev problem ({\it cf.\/} equation (1.9)),
\begin{eqnarray}
\ddot{x} - g \dot{y}\, + \epsilon  \gamma \dot{x} + \alpha x & = & 0
\nonumber\\
\ddot{y} + g \dot{x} \,+ \epsilon  \delta   \dot{y} + \beta  y & = & 0,
\end{eqnarray}
if $ \alpha $ and $ \beta $ are both negative, and if we set $ \epsilon \, = 0,
$ then for $ g^2 + \alpha + \beta > 2 \sqrt{\alpha \beta },$ all eigenvalues
are pure imaginary (the unperturbed system is gyroscopically stable). But, for
strong dissipation, $ \gamma > 0, \delta > 0 $ and $
\epsilon  \,> 0, $ one pair of eigenvalues crosses into the right half plane
and
another pair into the left half plane.  This was shown by a Routh-Hurwitz
calculation which, being a counting device, is not capable of telling us which
eigenvalue crosses over to which half plane.  Employing the crossing speed
formula (5.28) we are able to address precisely this problem of tracking
eigenvalue movement.

Note that the Chetaev problem is in the abelian case with
\begin{eqnarray}
M & = & \left( \begin{array}{cc}  1 & 0\\ 0 & 1\end{array}
\right);\, S = \left( \begin{array}{cc}  0 & -g\\ g & 0\end{array}
\right);\, R = \left( \begin{array}{cc}  \gamma  & 0 \nonumber\\ 0 & \delta
\end{array}  \right) \\
\Lambda & = & \left( \begin{array}{cc}  \alpha  & 0\\ 0 & \beta \end{array}
\right).
\end{eqnarray}
One checks that
\begin{equation}
{\rm det} ( G ( i \omega _0))   =  ( - \omega _0 ^2 + \alpha ) ( - \omega _0 ^2
+ \beta ) - \omega _0^2 g^2  =  0,
\end{equation}
which determines two distinct pairs of pure imaginary eigenvalues if $ g^2 +
\alpha + \beta > 2  \sqrt{\alpha \beta }. $ Corresponding to $ \lambda = i
\omega _0, $ a null-vector for $ G (i \omega _0) $ is given by,
\begin{equation} \left( \begin{array}{c}  q_1 \\ q_2\end{array}
\right) = \left( \begin{array}{c}  i g \omega_0\\ - \omega_0^2 + \alpha
\end{array}  \right) = \eta + i \beta.
\end{equation}
Thus $ \eta = ( 0, - \omega _0^2 + \alpha ) ^T, \, \, \beta = ( g \omega _0,
0)^T. $  Substituting in (5.28) we get,
\begin{equation}
\lambda ^\prime _r =   \frac{-\frac{1}{2}\{ \delta ( - \omega_0^2 + \alpha
)^2 + \gamma g^2 \omega_0^2\}}{\{ ( - \omega _0^2 + \alpha )^2 + g^2
\omega_0^2 + g^2 (- \omega_0^2 + \alpha )\}}.
\end{equation}
Using the relation (5.31) we can simplify further to obtain
\begin{equation}
\lambda ^\prime _r = - \frac{1}{2} \frac{\{ \delta ( - \omega_0^2 + \alpha
) + \gamma (- \omega_0^2 + \beta )\}}{\{ ( - \omega _0^2 + \alpha ) + (-
\omega_0^2 + \beta ) g^2\}}.
\end{equation}
Suppose $ \pm i \omega _0 $ and $ \pm i \omega_1 $ are the distinct eigenvalues
of the unperturbed system.  Then,
\[ \omega_0^2 + \omega_1^2 = g^2 + \alpha + \beta. \]
Therefore,
\[
(- \omega_0^2 + \alpha) + (- \omega_0^2 + \beta ) + g^2
= \omega_1^2 + \omega_0 ^2 - \omega_0 ^2 - \omega_0 ^2  =  \omega_1 ^2 - \omega
_0^2.\]
Thus,
\begin{equation}
\lambda ^\prime _r = \frac{ - \frac{1}{2} \{ \delta ( - \omega_0^2 + \alpha
) + \gamma (- \omega_0^2 + \beta )\}}{\{  \omega _1^2-
\omega_0^2 \}}.
\end{equation}
By hypothesis, $ \alpha < 0, \beta < 0, \delta > 0, \gamma > 0. $
It follows that
\[ -
\frac{1}{2} \{ \delta (-\omega_0^2 + \alpha ) + \gamma ( - \omega_0^2 + \beta )
\} > 0. \]
 Hence the simple eigenvalue $ \pm i\omega  _0 $ moves to the right
(left) half plane according as whether $ \omega _1 > \omega_0 ( \omega_1 <
\omega_0). $ See Figure 5.1. \quad $\blacklozenge$
\vspace{0.2in}

\begin{figure}%notes
\vspace{2.1in}
\hspace*{.8in} \special{illustration fig5.1.eps}
\caption{The  weaker get destabilized.}
\end{figure}

\noindent{\bf Example} The simplest non-abelian case arises when $G = SO(3)$
and
the shape space dimension is 1.  A physical example of this is that of a
rigid body with an
attached pointmass at the end of a spring, free to oscillate along a linear
guideway.  First, note that we can do some basic calculations without
reference to a particular equilibrium about which block-diagonal normal form
is used.  Let,
\[ L_\mu = \left[ \begin{array}{cc}  0 & -g\\ g & 0\end{array}
\right];\, C = \left[ \begin{array}{c} C_1\\C_2\end{array}
\right]; \,A_\mu = \left[ \begin{array}{cc} a_{11} & a_{12}\\ a_{12} &
a_{22}\end{array}  \right] > 0. \]
Note that $ C^T $ is not injective, a case not covered by Theorem {\bf 3.3}.
The magnetic term $ S  =  - S^T = 0, $ since the shape space dimension is $ 1 $
by hypothesis.  Let $ \Lambda = \alpha $ and $ M = m $ be the scalar stiffness
and mass respectively.  The quadratic pencil of Lemma {\bf 5.5} takes the
form
\[
G(\lambda) = \left[ \begin{array}{c c c} \lambda^2 m + \alpha  &  - \lambda C_1
 &  - \lambda C_2  \\
\\
 - \lambda C_2/g & \lambda -
\displaystyle{\frac{a_{12}}{g}}  & \displaystyle{- \frac{a_{22}}{g}} \\
\\
 \lambda
C_1/g &  \displaystyle{\frac{a_{11}}{g}} & \lambda + \displaystyle{
\frac{a_{12}}{g}} \end{array} \right]. \]
It can be verified that
\[ p(\lambda) = {\rm det}\, G(\lambda)
 =  m \lambda  ^4 + \lambda ^2 \left\{ \alpha + \frac{m}{g^2} \Delta _1 +
\frac{\Delta _2}{g^2} \right\} + \alpha \frac{\Delta _1}{g^2}, \]
where
\[ \Delta _1 = \frac{a_{11} a_{22} - a_{12}^2}{g^2} > 0 \]
(because $ A_\mu > 0) $ and
\begin{eqnarray*}
 \Delta _2& =& \frac{C_1^2 a_{22} + C_2^2 a_{11} - 2 C_1 C_2 a_{12}}{g^2} \\
& = & \frac{1}{g^2} (C_2 - C_1 )
\left[ \begin{array}{c c c} a_{11} & a_{12} \\ a_{12}& a_{22}
\end{array} \right]  \left( \begin{array}{c} C_2 \\ -C_1 \end{array} \right) >
0
\end{eqnarray*}
(again because $ A_\mu > 0). $

There are three cases to consider:
\begin{description}
\item [(a)] If $ \alpha > 0, $ then the second variation is postive
definite and all the
roots of $ p(\lambda), $ ({\it i.e.,\/} eigenvalues of the unperturbed
Hamiltonian system)
are pure imaginary.

\item [(b)] If $ \alpha = 0, $ then there is repeated root at the origin and a
pure imaginary pair $ \pm i \omega _0. $

\item [(c)]  If $ \alpha < 0, $ two of the roots of $ p(\lambda) $ are real
with
one root lying in the right half plane.
\end{description}

In case {\bf (a)}, a dissipative perturbation moves the eigenvalues into
the left
half plane.  This is already covered by the general theory, but can be
recovered by the crossing-speed formula (5.27) with $ S = 0, $ a calculation
left to the reader.  Case {\bf (c)} is the odd-index case and the
Cartan-Chetaev-Oh lemma demonstrates instability with or without added
dissipation. In case {\bf (b)} our crossing speed formula (5.27) applies to
the pair of pure imaginary roots (since they are simple).  The details are
again
left to the reader. \quad $\blacklozenge$

\section{Examples}
\noindent {\bf Example 1} (The Rigid Body with Internal Rotors)  Consider a
rigid body with
{\it two\/} symmetric rotors.  It is assumed that the rotors are subject to a
dissipative/frictional torque and no other forcing. A steady spin about the
{\it  minor axis\/} of the locked inertia tensor ellipsoid ({\it i.e.,\/}
the long
axis of the body), is a relative equilibrium.  Without friction, this system
can
experience gyroscopic stabilization and the second variation of the augmented
Hamiltonian can be indefinite. We aim to show that this is an unstable relative
equilibrium with dissipation added.

The equations of motion are (see Krishnaprasad [1985] and Bloch,
Krishnaprasad, Marsden, and Sanchez de Alvarez [1992]):
\begin{equation}\label{eqnofmotion}
\left. \begin{array} {l}
({\Bbb I}_{{\rm lock}} - {\Bbb I} _{{\rm rotor}}) \dot{\Omega }  =  ( {\Bbb
I}_{{\rm lock}} \Omega + {\Bbb I}_{{\rm rotor}} {\Omega } _r) \times
{\Omega } \\
\\
 \dot{\Omega
}_r  =  - ( {\Bbb I}_{{\rm lock}} - {\Bbb I}_{{\rm rotor}} )^{-1} ({\Bbb
I}_{{\rm lock}} {\Omega } + {\Bbb I}_{{\rm rotor}} {\Omega } _r)\times
\,{\Omega} - R {\Omega } _r \\
\\
 \dot{A}  =  A \hat{\Omega } \\
\\
\dot{\theta}_r  =  \Omega_r.
\end{array} \right\} \end{equation}

\vspace{0.2in}

\noindent In this example, $ Q = SO(3) \times S ^1 \times S ^1 $ and $ G =
SO(3) $. Also $ A \in
SO(3)  $ denotes the attitude/orientation of the {\it carrier\/} rigid body
relative to an inertial
frame, $ {\Omega } \in {\Bbb R}^3 $ is the body angular velocity of the
carrier, $ {\Omega } _r \in
{\Bbb R}^3 $ is the vector of angular velocities of the rotors in the body
frame (with
third component set equal to zero) and $ \theta_r $ is the ordered set of
rotor angles in
body frame (again, with third component set equal to zero).  Further, $
{\Bbb I}_{{\rm
lock}}$ denotes the moment of inertia of the body and locked rotors in the
body frame and
$ {\Bbb I}_{{\rm rotor}} $ is the 3 $ \times $ 3
 diagonal matrix of rotor inertias.  We let
\begin{equation}
\left.\begin{array}{rcl}
{\Bbb I}_{{\rm lock}} & = & {\rm diag} (B_1, B_2, B_3 ), \\
\\
{\Bbb I}_{{\rm rotor}} & = & {\rm diag} (J^1_1, J^2_2, 0), \\
\\
{\Bbb I}_{{\rm lock}} - {\Bbb I}_{{\rm rotor}} & = & {\rm diag} (A_1, A_2,
A_3).
\end{array} \right\} \end{equation}
Assume that $ B_1 > B_2 > B_3.$  Finally, $ R = {\rm diag} (R_1, R_2, 0) $
is the
matrix of {\it rotor\/} dissipation coefficients, $ R_i > 0. $

Consider the relative equilibrium for~(\ref{eqnofmotion}) defined by, $
{\Omega } ^e = (0, 0,
\omega )^T;\, \Omega ^e_r = (0, 0, 0)^T $ and $ \theta_r=\theta ^e_r $ an
arbitrary
constant.  This corresponds to a steady minor axis spin of the rigid body with
the two rotors non-spinning. Linearization of the $SO(3)  $-reduction
of~(\ref{eqnofmotion})
about this equilibrium yields,
\begin{equation}\label{lineqn}
\left.\begin{array}{l}
({\Bbb I}_{{\rm lock}} - {\Bbb I}_{{\rm rotor}}) \delta \dot{\Omega }  =  (
{\Bbb I}_{{\rm lock}} \delta \Omega + {\Bbb I}_{{\rm rotor}} \delta \Omega _r)
\times \Omega ^e  + \,({\Bbb I}_{{\rm lock}} \Omega ^e)
 \times \delta \Omega \nonumber \\
\\
\delta \dot{\Omega }_r  =  - ({\Bbb I}_{{\rm lock}} - {\Bbb I}_{{\rm
rotor}})^{-1} \left[ ({\Bbb I}_{{\rm lock}} \delta \Omega + {\Bbb I}_{{\rm
rotor}} \delta \Omega _r )  \times\, \Omega ^e \right. \nonumber \\
 \quad \quad \quad + \left.({\Bbb I}_{{\rm lock}} \Omega ^e)
\times \Omega \right]  - R \delta \Omega _r \nonumber\\
\\
\delta \dot{\theta}_r  =  \delta \Omega_r. \end{array} \right\} \end{equation}
It is easy to verify
that $ \delta \dot{\Omega }_3 = 0. $ This reflects the choice of relative
equilibrium. Similarly $ \delta \dot{\Omega
}_{r_{3}} = 0. $  We will now apply Theorem {\bf 3.3} in the case of $
\Lambda = 0
$.

Dropping the kinematic equations for $ \delta \theta _r $ we have the
``reduced'' linearized equations

\begin{equation}\label{redlineqn}
\left[ \begin{array}{c} \delta \dot{\Omega }_{r_{1}} \\
\\
 \delta
\dot{\Omega}_{r_{2}}\\
\\
 \delta \dot{\Omega}_1 \\
\\
 \delta \dot{\Omega}_2 \end{array} \right]
=
 \left[ \begin{array}{cccc} -R_1 & \displaystyle{\frac{-J_2^2  \omega
}{A_1}} &0  & \displaystyle{\frac{B_3 - B_2}{A_1}}\omega   \\
\\  \displaystyle{\frac{ J_1^1\omega  }{A_2}} & -R_2 & \displaystyle{\frac{B_1
-
B_3}{A_2}}\omega   & 0 \\
\\ 0 &  \displaystyle{\frac{J_2^2 \omega}{A_1}} & 0 & \displaystyle{\frac{B_2 -
B_3}{A_1}} \omega \\
\\
 \displaystyle{\frac{-J_1^1 \omega}{  A_2}} & 0&  \displaystyle{\frac{B_3 -
B_1}{A_2}}\omega &  0  \end{array} \right] \\
  \left[ \begin{array}{c} \delta \Omega _{r_{1}}\\
\\
 \delta \Omega
_{r_{2}} \\
\\
 \delta \Omega _1 \\
\\
 \delta \Omega _2
\end{array} \right].\end{equation}

\noindent Assume that $ \omega \neq 0 $ (nondegeneracy of the relative
equilibrium).  Then the above equations are easily verified to be in the normal
form (3.18), upon making the identifications, $ p = (\delta \Omega _{r_{1}},
\delta \Omega _{r_{2}})^T, q  =  (\delta \Omega _1, \delta
 \Omega _2)^T, \,\,\mbox{and},$
\[\begin{array}{rcl}
L_\mu  & = & \left( \begin{array}{cc} 0 & - 1/\omega \\ 1/\omega & 0
\end{array}
\right);\,\, C = \left( \begin{array}{cc} -1 &0 \\ 0 & -1 \end{array}
\right) ;\,\, \tilde{S} = \left( \begin{array}{cc} 0 & \omega \\ -\omega & 0
\end{array} \right)\\
\\
A_\mu & = & \left( \begin{array}{cc} \displaystyle{\frac{B_3 - B_1}{A_2}}
& 0 \\
0 & \displaystyle{\frac{B_3 - B_2}{A_1}}  \end{array} \right);\,\, M^{-1} =
\left( \begin{array}{cc}
\displaystyle{\frac{ J_1^1}{A_2}} & 0 \\ 0 & \displaystyle{\frac{J_2^2}{A_1}}
\end{array}
\right); \\
\\
R & = & \left( \begin{array}{cc} R_1 \displaystyle{\frac{A_2}{J_1^1}}  & 0
\\ 0 &
R_2 \displaystyle{\frac{A_1}{J_2^2}}  \end{array} \right).
\end{array}\]
Since $ B_1 > B_2 > B_3,\, A_\mu $ is negative definite. Also, $ M $ and $
R $ are
positive definite, and $ C^T $ is injective and thus all the hypotheses of
Theorem {\bf 3.3} are  satisfied.  Thus the linearized system~(\ref{lineqn})
or~(\ref{redlineqn}) displays dissipation-induced instability. \quad
$\blacksquare$
\vspace{0.2in}

\noindent{\bf Remark}\quad In the body and rotors example, the linearized
system
was shown to be in block-diagonal normal form by inspection.  Our calculations
also reveal that there is some freedom in the choice of block-diagonal
parameters-for instance the scalar $ \omega $ could appear in various ways in $
L_\mu, C $ etc.
\vspace{0.2in}

\noindent{\bf Remark}\quad This example is also instructive in that we can
verify the
instability result by a Routh-Hurwitz computation, as in Proposition {\bf
1.2}. We
sketch the computation here and note that calculations like this can
sometimes be
tedious, indicating the usefulness of the general result, even in this
relatively low
dimensional case.

 A straightforward calculation yields the following
characteristic polynomial of the linearized system (6.3) or (6.4), with {\it
no dissipation, \/}
\begin{equation}
p(\lambda) = \lambda^4 + \omega ^2 \lambda^2 \frac{(J_2^2 - (B_2
-B_3))(J_1^1 - (B_1-B_3))}{A_1 A_2}.
\end{equation}
There are two eigenvalues at the origin, consistent with the rank deficit of 2
in $\left( \begin{array}{cc} L^{-T}_\mu  & L_\mu ^{-T}C \\ C^T
L^{-T}_\mu  & -\tilde{S} \end{array} \right) $ and, under the additional
physical assumption that
\begin{equation}
(J_2^2 - (B_2 - B_3))(J_1^1 -(B_1 - B_3)) > 0,
\end{equation}
the other two eigenvalues are pure imaginary.
In fact, we assume both factors in the preceding equation (6.6) are
negative, since
the rotor inertias are small. Now consider the case in which $ R _1, R _2 >
0 $, and
are small. The full characteristic polynomial is \begin{eqnarray}
\lefteqn{\lambda ^4
+ \lambda ^3 (R _1 + R _2)} \nonumber \\  & & \mbox{} + \omega ^2 \lambda
^2 \left\{
\frac{ R _1 R _2 }{ \omega ^2} + \frac{ (J ^2 _2 - (B _2 - B _3))(J ^1 _1 -
(B _1 - B
_2))}{ A _1 A _2} \right\} \nonumber \\ & & \mbox{} - \omega ^2 \frac{
\lambda }{ A _1
A _2} \left\{ R _1 (B _1 - B _3)(J ^2 _2 - (B _2 - B _3)) + R _2 (B _2 - B
_3)(J ^1 _1
- (B _1 - B _3)) \right\} \nonumber \\ & & \mbox{} + \omega ^2 \left\{
\frac{(B _2 - B
_3)(B _1 - B _3) R _1 R _2 }{A _1 A _2} \right\}.
\end{eqnarray}
Now, use the same notation for the characteristic polynomial (6.7) as in
Proposition
{\bf 1.2}. We need to compute the sign changes in the sequence (1.14). Clearly
$\rho_1$ and $\rho_4$ are positive since $R _1$ and $R _2$ are positive and
$B_1 > B
_2 > B _3 $.

A computation shows that
\begin{eqnarray}
\rho_1 \rho_2 - \rho_3 & = & (R _1 + R _2)(R _1 R _2) + \frac{ \omega ^2 }{
A _1 A _2} J ^2 _2 J ^1 _1
\nonumber \\
& & \mbox{} - \frac{ \omega ^2 }{ A _1 A _2} \left\{ (B _1 - B _3)R _2 J ^2
_2 + (B _2 - B _3)J ^1
_1 R _1 \right\}.
\end{eqnarray}
The first two terms are small by assumption and hence $ \rho_1 \rho_2 -
\rho_3 $ is negative.
It then follows that \[ \frac{ \rho_3  (\rho_1 \rho_2 - \rho_3) - \rho_1
\rho_4 }{
\rho_1 \rho_2 - \rho_3} \] is positive.

Hence the Routh-Hurwitz sign sequence is $ \{ +, +, -, +, + \} $ and thus the
addition of dissipation has indeed moved two eigenvalues into the right
half plane,
causing a linear instability. \quad $\blacklozenge$  \vspace{0.2in}

\noindent{\large \bf Example 2} (Double Spherical Pendulum) In Marsden and
Scheurle
[1992] the double spherical pendulum is discussed. In particular, relative
equilibria,
called ``cowboy solutions'' are found explicitly and have a shape in which the
horizontal projections of the two rods point in opposite directions. The
group in this
case is $ S ^1 $, corresponding to rotations about the vertical axis. It is
verified
that indeed the linearized equations are in our standard form $ M \ddot{ q} + S
\dot{q} + \Lambda q = 0 $, but where the $ 3 \times 3 $ matrices $ M, S $ and
$\Lambda$ have extra zeros due to discrete symmetries. It is found that in
large
regions of parameter space (determined by the pendulum lengths, masses and
angular
momentum), that $\Lambda$ has signature $ (+, -, -) $, while the
eigenvalues of the
linearized system lie on the imaginary axis. It follows from Theorem {\bf
1.1} or {\bf
4.2} that if one adds joint friction (so that the total angular momentum is
still
conserved) then the cowboy solutions become spectrally unstable. This
example is a good
one in that direct analytical computation of eigenvalue movement to see this
instability would be quite complicated. We also point out that experiments of
John
Baillieul (Boston University) confirm this instability. \quad $\blacksquare$
\vspace{0.2in}

We also point out that similar eigenvalue and energetic situations arise in
a number
of other examples; among them are: \begin{enumerate}  \item The heavy top---
see
Lewis, Ratiu, Simo and Marsden [1992] \item The rotating liquid drop--- see
Lewis
[1989] \item Shear flow in a stratified fluid with Richardson number
between $ 1/4 $
and 1; see Abarbanel et al. [1986] \item Plasma dynamics; see Morrison and
Kotschenreuther [1989], Kandrup [1991], and Kandrup and Morrison [1992].
\end{enumerate}
The last three examples mentioned are infinite dimensional, which provide
motivation
for extending our methods to cover such cases. One infinite dimensional
example we can
handle is the next one.
\vspace{0.2in}

\noindent{\large \bf Example 3} We now consider a partial differential
equation for
which one can analyze dissipation induced instability by finite-dimensional
techniques.
We consider a Lagrangian for a model of a nonplanar rotating beam with
``square''
cross-section. The beam is assumed to be of Euler-Bernoulli type. It is
fixed to the
center of a circular plate rotating with constant angular velocity
$\omega$, with
undeflected position perpendicular to the plate along the $z$-axis of a
Cartesian
coordinate system fixed in the plate. The beam is inextensible and can
deflect in the
$x$- and $y$-directions. (The planar version of this model is analyzed in
Baillieul
and Levi [1987].) The Lagrangian is chosen to be
\begin{eqnarray}  L(x,y, x _t, y _t) & = &
\frac{1}{2} \int^1_0 \left( (x _t - y \omega) ^2 + (y _t + x \omega) ^2
\right) d z
\nonumber \\ & & \mbox{} - \frac{1}{2} \int^1_0 k (y^2 _{ zz} + x^2 _{ zz}
) d z .
\end{eqnarray}  where $k$ is an elastic constant.

The equations of motion with Rayleigh damping and damping constant $\gamma$
are:
\begin{equation}
\begin{array}{l}
x _{ tt} - 2 \omega y _t - \omega ^2 x + k x _{ zzzz} + \gamma x _t = 0 \\
y _{ tt} + 2 \omega x _t - \omega ^2 y + k y _{ zzzz} + \gamma y _t = 0 .\\
\end{array}
\end{equation}
The natural boundary conditions are:
\begin{equation}
\begin{array}{l}
x (0) = x' (0) = x'' (1) = x ''' (1) = 0 \\
y (0) = y' (0) = y'' (1) = y ''' (1) = 0 \\
\end{array}
\end{equation}
where ${}^\prime $ denotes the $z$-derivative.

The equilibrium states are given by $ x _{ tt}= y _{ tt} = 0, x _t = y _t =
0 $ and
hence \begin{equation}
\begin{array}{l}
- \omega ^2 x + k x _{ zzzz} = 0 \\
- \omega ^2 y + k y _{ zzzz} = 0 .
\end{array}
\end{equation}
We set $ k = 1 $ for convenience.

The fourth order operator with the given boundary conditions has compact
inverse and
hence (see e.g. Baillieul and Levi [1987]) the eigenvalues of equations
(6.9) are given
by $ 0 \leq \omega_1 ^2 \leq \omega_2 ^2 \leq \ldots \rightarrow \infty $ with
corresponding eigenfunctions $ x (z) = x _i (z) $ and $ y (z) = y _i (z) $
respectively. By our choice of the elastic constants, $ x_i (z) = y _i (z) $.

Consider now the undamped case, $ \gamma = 0 $, and write the equations in
first
order form, letting $ q _1 = x, q _2 = y, p _1 = x _t, p _2 = y _t $. We
obtain:
\begin{equation}  \begin{array}{l}  q _{ 1 t} = p _1 \\
q _{ 2 t} = p _2 \\
p _{ 1 t} = 2 \omega p _2 + \omega ^2 q _1 - q _{ 1 zzzz} \\
p _{ 2 t} = - 2 \omega p _1 + \omega ^2 q _2 - q _{ 2 zzzz} .\\
\end{array}
\end{equation}
Let $ z = [ q_1 \,  q_2 \,  p _1 \,  p _2 ]^T $. The system is thus of the
form $ z _t = A z $
where \begin{equation}
A = \left[ \begin{array}{c c c c} 0 & 0 & 1 & 0 \\ 0 & 0 & 0 & 1 \\ -
\partial ^4 + \omega ^2  & 0
& 0 & 2 \omega \\ 0 & - \partial ^4 + \omega ^2 & - 2 \omega & 0
\end{array} \right] .
\end{equation}
The stability of the equilibria are determined by the eigenvalues of $A$.
In addition to the
zero eigenvalue, one can check that $A$ has eigenvalues $ \pm i (\omega \pm
\omega _j )$ with
corresponding eigenvectors
\begin{eqnarray}
\left[ \begin{array}{l} x _j  \\ i x _j \\ i (\omega  + \omega _j)x _j \\ -
i (\omega  + \omega
_j) x _j  \end{array} \right]   & &
\left[ \begin{array}{l} x _j  \\ - i x _j \\ - i (\omega + \omega  _j)x _j \\ -
(\omega  + \omega
_j) x _j  \end{array} \right] \nonumber \\
\left[ \begin{array}{l} x _j  \\ i x _j \\ i (\omega _j  - \omega )x _j \\
- (\omega _j - \omega )
x _j  \end{array} \right] & &
\left[ \begin{array}{l} x _j  \\ - i x _j \\ - i (\omega _j  - \omega )x _j
\\ - (\omega _j -
\omega  ) x _j  \end{array} \right] . \end{eqnarray}

Now project the system onto the invariant subspace spanned by the four
eigenvectors corresponding
to the eigenvalues $ \pm i (\omega \pm \omega _j ) $. We see that on this
subspace we have a
gyroscopic system in Chetaev normal form (1.9). In fact, it is identical to
the system describing
the rotating bead given in \S {\bf 1}, with spring constant $ k = w_j ^2 $.
Hence for $ w ^2 > w
_1 ^2 $ we can see that addition of dissipation causes the system to become
spectrally unstable.
In fact, for $ w _j ^2 < w ^2 < w _j ^2 + 1 $ there are $j$ gyroscopically
stable Chetaev
subsystems whose eigenvalues will be driven into the right-half-plane on
the addition of
dissipation. \quad $\blacksquare$

\section*{References}

\begin{description}

\item Abarbanel, H.D.I., D.D. Holm, J.E. Marsden and T.S. Ratiu [1986]
Nonlinear stability analysis of stratified fluid equilibria,
{\it Phil. Trans. R. Soc. Lond. A\/} {\bf 318}, 349--409; also {\it Phys. Rev.
Lett.\/} {\bf 52} [1984] 2352--2355.

\item Abraham, R. and J.E. Marsden [1978]
{\it Foundations of Mechanics.\/} Second Edition,
Addison-Wesley Publishing Co., Reading, Mass..

\item Abraham, R., J.E. Marsden and T.S. Ratiu [1988]
{\it Manifolds, Tensor Analysis, and Applications.\/}
Second Edition, Springer-Verlag, New York.

\item Arnold, V. [1987]
{\it Dynamical Systems III.\/} Encyclopedia of Mathematics, Springer-Verlag.

\item Baillieul, J. and M. Levi [1987]
Rotational elastic dynamics,
{\it Physica D\/} {\bf 27}, 43--62.

\item Baillieul, J. and M. Levi [1991]
Constrained relative motions in rotational mechanics,
{\it Arch. Rat. Mech. An.\/} {\bf 115}, 101--135.

\item Bellman, R. [1963]
{\it Matrix Analysis.\/} Academic Press, New York.

\item Bloch, A.M., P.S. Krishnaprasad, J.E. Marsden and
T.S. Ratiu [1991]
Asymptotic stability, instability, and stabilization of relative equilibria,
{\it Proc. of ACC., Boston IEEE\/}, 1120--1125.

\item Bloch, A.M., P.S. Krishnaprasad, J.E. Marsden and
G. S\'{a}nchez de Alvarez [1992]
Stabilization of rigid body dynamics by internal and external torques,
{\it Automatica\/} {\bf 28}, 745--756.

\item Brockett, R. [1970]
{\it Finite Dimensional Linear Systems.\/} Wiley.

\item Chandrasekhar, K. [1977]
{\it Ellipsoidal Figures of Equilibrium.\/}
Dover.

\item Chetaev, N.G. [1961]
{\it The stability of Motion.\/}
Trans. by M. Nadler, Pergamon Press, New York.

\item Gantmacher, F.R. [1959]
{\it Theory of Matrices.\/}
Chelsea, N.Y.

\item Guckenheimer, J. and A. Mahalov [1992],
Instability induced by symmetry reduction, {\it Phys. Rev. Lett., \/} {\bf 68},
2257--2260.

\item Hahn, W. [1967]
{\it Stability of Motion.\/}
Springer-Verlag, New York.

\item Haller, G. [1992]
Gyroscopic stability and its loss in systems with two essential coordinates,
{\it Int. J. Nonlinear Mech.\/} {\bf 27}, 113--127.

\item Holm, D.D., J.E. Marsden, T.S. Ratiu and A. Weinstein [1985]
Nonlinear stability of fluid and plasma equilibria,
{\it Phys. Rep.\/} {\bf 123}, 1--116.

\item Kandrup,  H.E. [1991]
The secular instability of axisymmetric collisionless star cluster,
{\it Astrophy.  J.\/} {\bf 380}, 511--514.

\item Kandrup,  H.E. and P. Morrison [1992]
Hamiltonian structure of the Vlasov-Einstein system and the problem of
stability
for spherical relativistic star clusters,
{\it preprint.\/}.

\item Krein, M.G. [1950]
A generalization of some investigations of linear differential equations
with periodic
coefficients. {\it Doklady Akad. Nauk SSSR N.S.\/} {\bf 73}, 445--448.

\item Krishnaprasad, P.S. [1985]
Lie-Poisson structures, dual-spin spacecraft and
asymptotic stability, {\it Nonl. An. Th. Meth. and Appl.\/} {\bf 9},
1011--1035.

\item Krishnaprasad, P.S. and J.E. Marsden [1987]
Hamiltonian structure and stability for rigid bodies with flexible attachments,
{\it Arch. Rat. Mech. An.\/} {\bf 98}, 137--158.

\item LaSalle, J.P. and S. Lefschetz [1963]
{\it Stability by Lyapunov's direct method.\/}
Academic Press, New York.

\item Levi, M. [1977]
Stability of linear Hamiltonian systems with periodic coefficients.
{\it Research Report\/} RC 6610(\#28482), IBM F.J. Watson Research Center.

\item Lewis, D.K. [1989]
Nonlinear stability of a rotating planar liquid drop,
{\it Arch. Rat. Mech. Anal.\/} {\bf 106}, 287--333.

\item Lewis, D.K. [1992] Lagrangian
block diagonalization. {\it Dyn. Diff. Eqn's.\/} {\bf 4} 1--42.

\item Lewis, D.K. [1993]
Linearized dynamics of symmetric Lagrangian systems, to appear in
{\it International Conference of Hamiltonian Dynamical Systems,\/} Cincinnati,
March 25--28, 1992.

\item Lewis, D.K. and T.S. Ratiu [1993] (manuscript in preparation).

\item Lewis, D., T.S. Ratiu, J.C. Simo and J.E. Marsden [1992]
The heavy top, a geometric treatment,
{\it Nonlinearity\/} {\bf 5}, 1--48.

\item Lewis, D.K. and J.C. Simo [1990]
Nonlinear stability of rotating pseudo-rigid
bodies, {\it Proc. Roy. Soc. Lon. A\/} {\bf 427}, 281--319.

\item MacKay, R. [1991]
Movement of eigenvalues of Hamiltonian equilibria under non-Hamiltonian
perturbation, {\it Phys. Lett.\/} {\bf A155}, 266--268.

\item Marsden, J.E. and J. Scheurle [1992]
Lagrangian reduction and the double spherical pendulum,
(ZAMP, to appear).

\item Marsden, J.E., J.C. Simo, D.K. Lewis and T.A. Posbergh [1989]
A block diagonalization theorem in the energy momentum method,
{\it Cont. Math. AMS\/} {\bf 97}, 297--313.

\item Morrison, P.J. and M. Kotschenreuther [1989] The free energy principle,
negative energy modes and stability, {\it Proc 4th Int. Workshop on
Nonlinear and
Turbulent Processes in Physics,\/} World Scientific Press.

\item Oh, Y.G., N. Sreenath, P.S. Krishnaprasad and J.E. Marsden [1989]
The Dynamics of Coupled Planar Rigid Bodies Part 2:
Bifurcations, Periodic Solutions, and
Chaos, {\it Dynamics and Diff. Eq'ns.\/} {\bf 1}, 269--298.

\item O'Reilly, O. [1993]
Reversible dynamical systems, destabilization, and follower force problems,
{\it preprint.\/}

\item Oh, Y.-G. [1987]
A stability criterion for Hamiltonian systems with symmetry,
{\it J. Geom. Phys.\/} {\bf 4}, 163--182.

\item Poincar\'{e}, H. [1885]
Sur l'\'{e}quilibre d'une masse fluide anim\'{e}e d'un mouvement de rotation,
{\it Acta. Math.\/} {\bf 7}, 259.

\item Poincar\'{e}, H. [1892]
Les formes d'\'{e}quilibre d'une masse fluide en rotation,
{\it Revue G\'{e}n\'{e}rale des Sciences\/} {\bf 3}, 809--815.

\item Poincar\'{e}, H. [1901]
Sur la stabilit\'{e} de l'\'{e}quilibre des figures piriformes
affect\'{e}es par une masse
fluide en rotation,
{\it Philosophical Transactions A\/} {\bf 198}, 333--373.

\item Riemann, B. [1860]
Untersuchungen \"{u}ber die Bewegung eines fl\"{u}ssigen gleich-artigen
Ellipsoides, {\it Abh. d. K\"{o}nigl. Gesell. der Wiss. zu G\"{o}ttingen\/}
{\bf
9}, 3--36.

\item Routh, E.J. [1877]
{\it Stability of a given state of motion.\/}
Reprinted in Stability of Motion, ed. A.T. Fuller, Halsted Press, New York,
1975.

\item Simo, J.C., D.K. Lewis and J.E. Marsden [1991]
Stability of relative equilibria I: The reduced energy momentum method,
{\it Arch. Rat. Mech. Anal.\/} {\bf 115}, 15-59.

\item Simo, J.C., T.A. Posbergh and J.E. Marsden [1990]
Stability of coupled rigid body and geometrically exact rods:
block diagonalization and the energy-momentum method,
{\it Physics Reports\/} {\bf 193}, 280--360.

\item Simo, J.C., T.A. Posbergh and J.E. Marsden [1991]
Stability of relative equilibria II: Three dimensional elasticity,
{\it Arch. Rat. Mech. Anal.\/} {\bf 115}, 61--100.

\item Sri Namachchivaya, N., S.T. Ariaratnam [1985]
On the dynamic stability of gyroscopic systems,
{\it SM Archives\/} {\bf 10}, 313--355.

\item Taussky, O. [1961] A Generalization of a Theorem of Lyapunov
{\it SIAM J. Appl. Math\/} {\bf 9}, 640--643.

\item Thomson, L. and P.G. Tait [1912]
{\it Principles of Mechanics and Dynamics.\/}
Cambridge Univ. Press (reprinted by Dover Publications Inc., 1962).

\item van Gils, S.A., M. Krupa and W.F. Langford [1990]
Hopf bifurcation with non-semisimple 1:1 resonance,
{\it Nonlinearity\/} {\bf 3}, 825--830.

\item Wang, L.S. and P.S. Krishnaprasad [1992]
Gyroscopic control and stabilization,
{\it J. Nonlinear Sci.\/} {\bf 2}, 367--415.

\item Whittaker, E.T. [1937]
{\it A Treatise on the Analytical Dynamics of Particles and Rigid Bodies.\/}
4th ed., Cambridge University Press. Reprinted, 1989.

\item Ziegler, H. [1956] On the concept of elastic stability, {\it Adv. Appl.
Mech.\/} {\bf 4}, 351--403.

\end{description}

\end{document}